\documentclass[pra,twocolumn,aps,showpacs,nofootinbib,floatfix,nobibnotes]{revtex4-2}
\usepackage{amsmath, bm}
\usepackage{amssymb}
\usepackage{graphicx}
\usepackage{verbatim}
\usepackage[colorlinks=true,linkcolor=blue,citecolor=blue,urlcolor=blue]{hyperref}
\usepackage{txfonts}
\usepackage{dsfont}
\usepackage{enumitem}
\usepackage[utf8]{inputenc}

\usepackage{hhline}
\usepackage{multirow}
\usepackage{makecell}

\usepackage{booktabs}
\begin{document}

\title{Two-photon absorption measurements in the presence of single-photon losses}
\author{Shahram Panahiyan$^{1,2}$, Carlos S\'anchez~Mu{\~n}oz$^3$, Maria V. Chekhova$^{4,5}$, and Frank Schlawin$^{1,2}$}
\email{frank.schlawin@mpsd.mpg.de}
\affiliation{$^1$ Max Planck Institute for the Structure and Dynamics of Matter, Luruper Chaussee 149, 22761 Hamburg, Germany }
\affiliation{$^2$ The Hamburg Centre for Ultrafast Imaging, Luruper Chaussee 149, Hamburg D-22761, Germany}
\affiliation{$^3$ Departamento de F\'isica Te\'orica de la Materia Condensada and Condensed Matter Physics Center (IFIMAC), Universidad Aut\'onoma de Madrid, Madrid, Spain}
\affiliation{$^4$ Max-Planck Institute for the Science of Light, Staudtstr. 2, Erlangen D-91058, Germany}
\affiliation{$^5$ University of Erlangen-Nuremberg, Staudtstr. 7/B2, Erlangen D-91058, Germany}

\date{\today}

\begin{abstract}

We discuss how two-photon absorption (TPA) of squeezed and coherent states of light can be detected in measurements of the transmitted light fields. 
Such measurements typically suffer from competing loss mechanisms such as experimental imperfections and linear scattering losses inside the sample itself, which can lead to incorrect assessments of the two-photon absorption cross section. 
We evaluate the sensitivity with which TPA can be detected and find that TPA sensitivity of squeezed vacua or squeezed coherent states can become independent of linear losses at sufficiently large photon numbers. In particular, this happens for measurements of the photon number or of the anti-squeezed field quadrature, where large fluctuations counteract and exactly cancel the degradation caused by single photon losses.

\end{abstract}

\maketitle

\section{Introduction}

Nonclassical quantum states of light are gaining prominence as resources for quantum-enhanced applications in imaging and spectroscopy. Of particular interest in this endeavour are nonlinear light-matter interactions, where the most stunning effects are predicted. 
Two-photon absorption of entangled photons or squeezed light holds great promise for imaging and spectroscopic applications, since, at low photon fluxes, the absorption probability scales linearly with the light field intensity~\cite{Klyshko1982, Gea89, Javainen90, Georgiades95, Georgiades97, Dayan04} - in contrast to the quadratic scaling of the TPA signal induced by laser light. This remarkable effect could enable nonlinear spectroscopy and microscopy at low photon fluxes, which will be highly beneficial for photosensitive samples. To date, TPA of entangled photons or squeezed states of light was reported both in atomic gases~\cite{Georgiades95, Dayan04, Dayan05} and in molecules \cite{Lee06, Guzman10, Upton13, Villabona17, Villabona18, Li20b, Tabakaev2021, Burdick2021}, even though substantial controversy persists regarding the achievable enhancement of TPA signals due to entanglement in the latter case~\cite{Landes2021, Landes2021b, Raymer2021, Parzuchowski2021}. Entangled TPA could form the basis of new spectroscopic and microscopy applications \cite{Dorfman16, JPhysB17, AccChemRes,Gilaberte2019,Szoke2020, Mukamel2020,Ma2021, Eshun2022}, where quantum correlations reveal otherwise hidden features or increase the signal-to-noise ratio.%, and give rise to the emerging field of quantum biophotonics. 

Beams of isolated entangled photon pairs, however, suffer inevitably from very low photon count rates, thus necessitating long measurement times, and rendering them less appealing for practical applications. It is therefore important to also investigate quantum spectroscopy at higher photon fluxes~\cite{Raymer2022}. 
At higher photon fluxes, entangled photon pairs start to overlap temporally to form squeezed vacuum states of light, which can contain macroscopic photon numbers~\cite{Bondani2007, Iskhakov16, Beltran2017, Chekhova2018, Sharapova2020, Florez2020}, 
and still show interesting properties for applications. 
%Squeezed states feature large photon number fluctuations which still contain interesting properties for applications. 
For instance, large photon number fluctuations render them more efficient at driving higher harmonic generation in crystals than laser pulses with identical intensity~\cite{Spasibko17}. 
These states could further provide benefits for quantum imaging~\cite{Brida2009, Brida2010, Lopaeva2013, deAndrade2020, Casacio2021, Varnavski2020, Moreau2019} or spectroscopy~\cite{Tapster91, Polzik92, Moreau2017, Li20, Birchall2020, Atkinson2021}, due to the reduced quadrature fluctuations of squeezed light sources. Moreover, spectral quantum correlations, which lie at the heart of quantum advantage in entangled two-photon absorption, also remain prominent in such higher-intensity states~\cite{Cutipa2022}, and could be exploited for sensing.
It is in this regime that recent experimental demonstrations of stimulated Raman imaging with squeezed light provided conclusive evidence that quantum properties of light can improve the performance of nonlinear imaging and spectroscopic applications beyond the limits of classical laser pulses~\cite{deAndrade2020, Casacio2021}. Quantum enhancement of the signal-to-noise ratio can reduce photodamage, which constitutes a major drawback of nonlinear methods in the investigation of photosensitive biological samples, and open novel perspectives in quantum-enabled imaging technologies.

%Nonlinear optics plays a crucial role in sensing applications~\cite{Boyd}, including basic science as well as metrology. In imaging applications, nonlinear excitation can be a powerful tool to beat the single-photon diffraction limit and enhance the resolution~\cite{So00, Hell2015}, using strong ultrafast laser sources. In the quantum regime, nonlinear optics offers novel opportunities for applications as well, such as the establishment of a new radiation standard~\cite{Lemieux2019}. 

Nonlinear spectroscopy with high-flux squeezed states of light was also discussed extensively in the theoretical literature already. 
The application to nonlinear spectroscopy was first investigated theoretically in \cite{Roslyak09}, where the distinct scaling behaviour of different matter pathways contributing to the signal was analyzed. The spectroscopic information contained in squeezed light emission was considered both experimentally and theoretically in \cite{Yang2020, Dorfman2021}.
More recent work by Michael et al. analyzed squeezing-enhanced detection of coherent Raman scattering in a nonlinear interferometer~\cite{Michael19, Michael2021}, demonstrating how interferometric measurements could be exploited in nonlinear spectroscopy. 
Finally, a recent analysis by some of the authors showed that squeezing can enhance the precision with which two-photon absorption losses could be measured \cite{Carlos2021}, based on calculations of classical and quantum Fisher information. In particular, the maximally achievable precision for determining TPA losses with quadrature measurements was shown to improve quartically with the mean photon photon number, $\propto n^{4}$. Such a scaling can only be achieved using squeezed states, while coherent (classical) states only enable a cubic scaling $\propto n^3$. In other observables such as the photon number or the anti-squeezed quadrature, squeezed states only realize a quadratic scaling, whereas coherent states again have a cubic dependence on the incident photon number.
These results were obtained, however, in an idealised situation where no competing loss sources impede the precise measurement of the TPA. As these will be unavoidable in any experiment, however, the role of these single-photon losses remains an important open question. 

In this paper, we will address it and investigate in detail the influence of single-photon losses on the measurement of the TPA absorbance of a sample. We will show that the quartic scaling mentioned above is quickly degraded even by very weak single-photon losses. 
In stark contrast, the sensitivity of observables that scale quadratically with the photon number becomes \textit{independent} of these losses.  
Measurements with coherent states are always affected by the single-photon losses. 
Combining these findings, we find that amplitude-squeezed states can eliminate the degradation of the sensitivity due to the single-photon losses in measurements. 
As a consequence, for instance, photon number measurements with sufficiently strongly amplitude-squeezed states can combine the superior cubic scaling of the measurement sensitivity with a robustness against degradation by noise. 
This insight allows us to derive an ``optimal degree of squeezing". It depends on the noise level, i.e. on the single-photon loss probability due to undesirable error sources, as well as on the mean  photon number of the incident photonic state.
% for any given noise level and mean photon number of the input state. 

The paper is structured as follows: 
In Section~\ref{sec.TPA} we introduce the model of TPA losses, which we describe with a Markov master equation. We describe how losses are modelled and how signals are calculated. 
In Section~\ref{sec.results} we apply this formalism to TPA measurements in transmission. We calculate the classical Fisher Information (FI) in the presence of losses, and investigate how the information gets eroded. 
Finally, in Section~\ref{sec.conclusions} we summarize our findings.

\section{Measuring Two-photon absorption}
\label{sec.TPA}

\subsection{Master equation}

We are interested in the transmission of a quantum state of light through a two-photon absorbing medium. 
To describe this situation, we eliminate the material using the normal methods of open quantum systems to obtain a Markovian Lindblad equation for the density matrix of the light field. In the rotating frame with respect to the field Hamiltonian it reads~\cite{Simaan75, Simaan75b, Simaan78, Gilles93}
\begin{align}
\frac{d}{dt} \rho &= \gamma_{TPA} \mathcal{L}_{TPA} \rho = \frac{\gamma_{TPA}}{4} \left( 2 a^2 \rho a^{\dagger 2} - a^{\dagger 2} a^2 \rho - \rho a^{\dagger 2}  a^2\right). \label{eq.Lindblad}
\end{align}
Here, $a$ ($a^\dagger$) denotes the photon annihilation (creation) operator of the field mode.
Note that, in writing Eq.~(\ref{eq.Lindblad}), we write down the Lindblad operator given by a correlated loss operator $L = a^2 / \sqrt{2}$, in order to simplify expressions in our subsequent derivations. In the remainder of this paper we will be interested in the measurement of the absorbance $\varepsilon \equiv \gamma_{TPA} t$, where $t$ is the propagation time of the light through the TPA medium.
%where $t = \ell /c$ is the propagation time of the light through the TPA medium. 

\subsection{Measurement sensitivity \& Cramer-Rao bounds}
\label{sec.sensitivity}
%We will be interested in the sensitivity with which the absorbance $\varepsilon$ can be measured.  
%Here we are interested in t
The sensitivity of the measurement of the absorbance $\varepsilon$ via the expectation value of a general operator $\mathcal{O}$, $\Delta\varepsilon_{\mathcal{O}}$, is defined as the square root of the variance $\Delta\varepsilon^2_{\mathcal{O}}$. It can be obtained from error propagation \cite{Braunstein94},
\begin{align}
\Delta\varepsilon_{\mathcal{O} }^2 &= \frac{ \text{ Var } (\mathcal{O}) }{ \left| \frac{\partial \langle \mathcal{O} \rangle}{ \partial \varepsilon } \right|^2}, \label{eq.sensitivity}
\end{align}
where the expectation value is evaluated at fixed $\varepsilon$. In our derivation below, we will evaluate the changes at $\varepsilon = 0$, i.e. we assume that we can approximate the transmitted density matrix as $\rho \simeq \rho_0 + \varepsilon \times(\partial\rho_0 / \partial \varepsilon)$. 
%This seems justified for most applications as typical two-photon absorption cross sections in molecules are of the order $1 - 1000$~GM where $1$~GM $= 10^{-58}$~m$^4$s \cite{Landes2021b}, and so TPA losses are very weak. 
This seems justified for most applications, as typical two-photon absorption cross sections in molecules are very small. 
It is this small-$\varepsilon$ limit that will be the focus of our attention. The generalization to finite $\varepsilon$ is conceptually straightforward, but does not allow analytical results and thus prevents us from developing an intuitive understanding of the underlying physics.

In general, the sensitivity~(\ref{eq.sensitivity}) is bounded by the classical FI $\mathcal{F}_{cl} (\mathcal{O}, \rho)$ associated with the measurement of $\mathcal{O}$, which in turn is bounded by the quantum Fisher information $\mathcal{F}_{Q} ( \rho)$, which is further maximised over any positive operator-valued measurement~\cite{Braunstein94, Paris09, Haase16},
\begin{align}
\Delta\varepsilon_{\mathcal{O} }^2 &\geq \frac{1}{ \mathcal{F}_{cl} (\mathcal{O}, \rho) } \geq  \frac{1}{ \mathcal{F}_{Q} ( \rho) }. \label{eq.Cramer-Rao}
\end{align}
These inequalities are known as the classical and quantum Cramer-Rao bounds, respectively. As shown in \cite{Carlos2021}, the quantum FI diverges in the limit of very small absorbances, $\varepsilon \rightarrow 0$ and consequently does not provide a useful tool to assess the achievable sensitivity. Instead, we will focus on the classical Fisher information in the following.

%The sensitivity of photon number measurements can be enhanced further by the analysis of the whole
The FI of a measurement such as the detection of photon numbers is a function of the probability distribution pertaining to this measurement. 
Here, the photon number distribution is given by the set of probabilities $\{ P_n \}$ for the detection of $n$ photons. The information to be gained from the change of this distribution under TPA losses is quantified by the FI \cite{Paris09, Haase16}
\begin{align}
\mathcal{F}_C (\rho_\varepsilon, \hat{n}) &= \sum_n P_n (\varepsilon) \left( \frac{\partial}{\partial \varepsilon} \ln P_n (\varepsilon) \right)^2 \notag \\
&= \sum_n \frac{1}{P_n (\varepsilon)} \left(\frac{\partial P_n (\varepsilon)}{ \partial \varepsilon }\right)^2, \label{eq.FI_PhotoNum}
\end{align}
It is likewise defined for operators with continuous spectrum, which give rise to a probability distribution function.
For instance, given the probability distribution $P(q)$ of possible measurement outcomes for measurements of the field quadrature $\hat{q}$, we calculate the corresponding FI
\begin{equation}
\mathcal{F}_C (\rho_\varepsilon, q) = \int \!\! dq \frac{1}{P (q)}\left(\frac{ dP(q)}{d\varepsilon }\right)^2. \label{eq.quad-CFI}
\end{equation}

\begin{figure*}[t]
\centering
\includegraphics[width=0.9\textwidth]{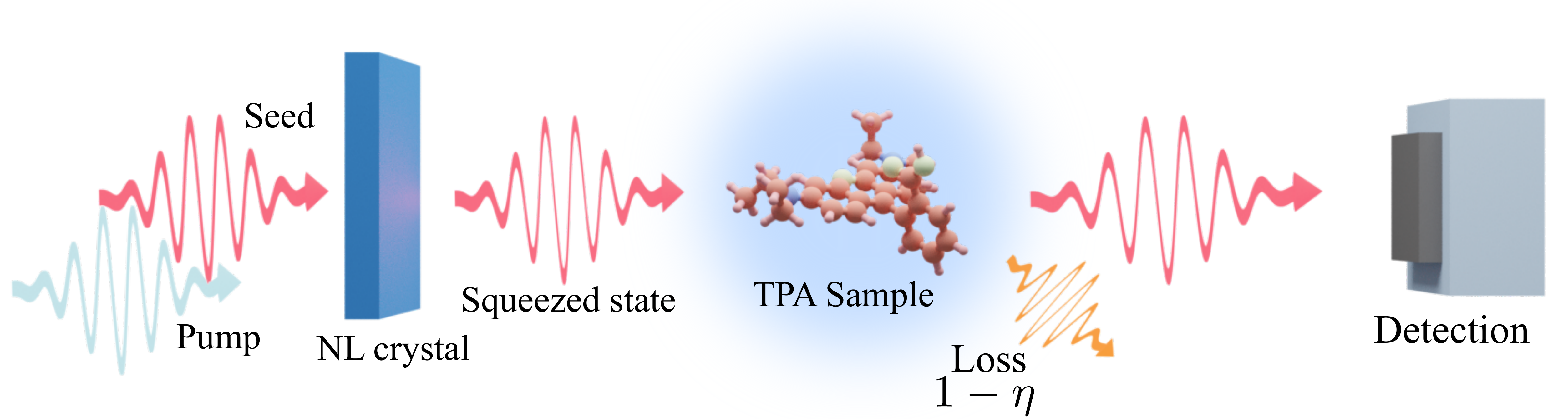}
\caption{
Two-photon absorption in a sample is detected by measuring the transmission of a coherent or squeezed beam. A pump pulse (blue) with frequency $2\omega_0$ drives downconversion into photons at $\omega_0$ in a nonlinear crystal creating a squeezed state or a squeezed coherent state in case the downconversion process is seeded (red). 
The output field is focused on the TPA sample and the transmitted light is detected. Single photon losses with a loss rate $1 - \eta$ can degrade the measurement outcome.
}
\label{fig.entTPA}
\end{figure*}

\subsection{Measurement sensitivity for two-photon losses}

We now discuss how we evaluate measurement observables and the measurement sensitivities. We consider TPA measurements in a transmission geometry, as sketched in Fig.~\ref{fig.entTPA}. An input coherent seed at frequency $\omega_0$ is squeezed in an optical parametric amplifier (OPA), implemented by means of a nonlinear crystal pumped at frequency $2\omega_0$, and then used as a probe for a TPA measurement. 
In this setup, the expectation value of the operator $\mathcal{O}$ is formally given by
\begin{align}
\langle \mathcal{O} \rangle &= \text{tr} \left\{ \mathcal{O} e^{ \mathcal{L}_{loss} } e^{ \mathcal{L}_{TPA} \varepsilon } e^{ \mathcal{L}_{OPA} }  \rho_0 \right\}, \label{eq.expectation-value}
\end{align}
where $\rho_0$ is the initial state of light. 
$\mathcal{L}_{OPA}$ is a superoperator, which describes a %carry out 
squeezing operation, i.e. it acts on a density matrix $\sigma$ as
\begin{align}
e^{ \mathcal{L}_{OPA} }  \sigma &\equiv U_{OPA} \sigma U^{\dagger}_{OPA}. \label{eq.U_OPA}
\end{align}
Here, we defined the squeezing transformation~\cite{BarnettRadmore} 
\begin{align}
U_{OPA} = \exp \left(  \frac{\zeta}{2} a^{\dagger 2} - \frac{\zeta^\ast}{2} a^2 \right), \label{eq.U_OPA}
\end{align}
and $\zeta = r e^{i\phi_r}$. %U_{OPA} = \exp (\zeta a^{\dagger 2}/2-\zeta^\ast a^2/2 )
The second superoperator in Eq.~(\ref{eq.expectation-value}), $\exp ( \mathcal{L}_{TPA} \varepsilon )$, describes the evolution according to the two-photon loss Lindbladian~(\ref{eq.Lindblad}).
Finally, we account for single-photon losses, which could stem from scattering losses in the optical system or imperfect photon detection. 
%We model these losses as scattering into an auxiliary photon mode described by the photon annihilation operator $c$, i.e. they are described by the superoperator
We can treat these losses via a beam splitter transformation in which the scattering into an auxiliary photon mode with photon annihilation operator $c$ is described by a superoperator
\begin{align}
e^{ \mathcal{L}_{loss} } \sigma &= U_{loss} \sigma U^{\dagger}_{loss},
\end{align}
where
\begin{align}
U_{loss} &= \exp \left(\tau (a c^\dagger + c a^\dagger) /2 \right), \label{eq.lin-loss}
\end{align}
and $\tau =  \arccos (\sqrt{\eta} )$. Here, $\eta$ denotes the transmission probability for a photon, and $1 - \eta$ quantifies the losses which are incurred by this process.

We can now calculate Eq.~(\ref{eq.sensitivity}). 
Since only the TPA evolution depends on $\varepsilon$, we find straightforwardly the change of the expectation value with $\varepsilon$,
\begin{align}
\frac{\partial \langle \mathcal{O} \rangle}{ \partial \varepsilon }  \bigg|_{\varepsilon = 0} &= \text{tr} \left\{ \mathcal{O} e^{\mathcal{L}_{loss} } \frac{ \partial }{ \partial \varepsilon} e^{ \mathcal{L}_{TPA} \varepsilon } e^{ \mathcal{L}_{OPA } }  \rho_0 \right\}  \bigg|_{\varepsilon = 0} \notag \\
&= \text{tr} \left\{ \mathcal{O} e^{\mathcal{L}_{loss} }  \mathcal{L}_{TPA} e^{ \mathcal{L}_{OPA } }  \rho_0 \right\}. \label{eq.derivative}
\end{align}
Using Eq.~(\ref{eq.U_OPA}), we can rewrite Eq.~(\ref{eq.derivative}) into an expectation value with respect to the initial state $\rho_0$. 
Defining the primed operators $\mathcal{O}' \equiv U^{\dagger}_{TPA} \mathcal{O} U_{TPA}$, we obtain
\begin{align}
\mathcal{O}_f &=  \mathcal{L'}_{TPA} [ U^{\dagger}_{loss} \mathcal{O}' U_{loss} ] ,
\end{align}
and 
\begin{align}
\frac{\partial \langle \mathcal{O} \rangle}{ \partial \varepsilon }  \bigg|_{\varepsilon = 0} &= \text{tr} \left\{ \mathcal{O}_f \rho_0 \right\}. \label{eq.O_f}
\end{align}
Here, $\mathcal{L'}_{TPA}$ denotes the adjoint superoperator to Eq.~(\ref{eq.Lindblad}), which acts on a operator $X$ as
\begin{align} \label{eq.L^adj}
\mathcal{L'}_{TPA} X &= \frac{1}{4} \left( 2 a'^{\dagger 2} X a'^2 - a'^{\dagger 2} a'^2 X - X a'^{\dagger 2} a'^2\right),
\end{align}
where we define the transformed operators $a' = U^\dagger_{OPA } a U_{OPA}$ and $a'^\dagger = U^\dagger_{OPA} a^\dagger U_{OPA}$.

Eq.~(\ref{eq.O_f}) can be translated into successive transformations of the photon operators in the Heisenberg picture, 
\begin{enumerate}[label=\arabic*)]
\item Perform squeezing operation. Using Eq.~(\ref{eq.U_OPA}), we find
%The squeezing transformation reads~\cite{BarnettRadmore} 
%\begin{align}
%U_{OPA} &= \exp \left( \frac{\xi}{2} a^2 - \frac{\xi^\ast}{2} a^{\dagger 2} \right),
%\end{align}
%with $\xi = r \exp ( i \phi_r)$, which yields the transformation
\begin{align}
a \rightarrow a' = U^\dagger_{OPA } a U_{OPA} = \cosh (r) a + \sinh (r) e^{i \phi_r} a^\dagger. \label{eq.OPA1}
\end{align}
%\textcolor{red}{Factor "-1" in Eq.~(\ref{eq.OPA1}) may be wrong.. $\phi_r$ or $\phi_r + \pi$?}
Unless specified otherwise, we will set $\phi_r = 0$ in the following without loss of generality.
\item Apply the adjoint Lindbladian~(\ref{eq.L^adj}) to $a'$, \\
\begin{align}
a' \rightarrow a'' &= \mathcal{L'}_{TPA} [a'] = - \frac{1}{2} a'^{\dagger} a'^{2}. \label{eq.L-trafo}
\end{align}
\item Perform the beamsplitter transformation according to Eq.~(\ref{eq.lin-loss}) to account for single photon losses, \\
\begin{align}
a'' \rightarrow a''' = \sqrt{\eta} a'' + \sqrt{1 - \eta} c. \label{eq.loss-trafo}
%a'' \rightarrow a''' = \cosh (r_2) a'' + e^{i \phi_2} \sinh (r_2) a^{''\dagger} \label{eq.squeezing-trafo}
\end{align}
\end{enumerate}
Finally, the expectation value is taken with respect to the initial state of the light field, which in the following we will assume to be either in the vacuum or in a coherent state with amplitude $\alpha = |\alpha|e^{i \phi}$.

\subsection{Calculation of the Fisher Information}

In contrast to calculating expectation values, the evaluation of the classical FI requires us to determine the full probability distribution $\{ P_n \}$ of measuring $n$ photons, or the continuous distribution functions $P (q)$ and $P (p)$ for the measurement of field quadratures, respectively. As described in \cite{Carlos2021}, this is best done in the Schr\"odinger picture in a squeezed basis defined by the orthonormal basis $\vert \tilde{n} \rangle = U_{OPA} \vert n \rangle$. In this basis, the squeezed vacuum state becomes the vacuum and a squeezed coherent state simply becomes a coherent state\footnote{We can also transform to a squeezed-displaced basis defined by $\vert \tilde{\tilde{n}} \rangle = U_{OPA} D(\alpha) \vert n \rangle$, where the squeezed coherent state again becomes the vacuum.}. The time evolution is then given by Eq.~(\ref{eq.Lindblad}), where the photon annihilation operators have to be transformed into the squeezed basis, 
\begin{align}
\tilde{a} = U^\dagger_{OPA} a U_{TPA}.
\end{align}
The change of the density matrix is then given by the action of the transformed Lindbladian on the ground state of the squeezed basis, i.e. $\partial \tilde{\rho} / \partial\varepsilon = \tilde{\mathcal{L}}_{TPA} \vert \tilde{0} \rangle \langle \tilde{0} \vert$. 

\section{Results}
\label{sec.results}

%In order to better understand the simulations of the SU (1,1) interferometer which we discuss later, we first consider the optical setup depicted in Fig.~\ref{fig.entTPA}, where the light field created by a downconversion source is transmitted through a TPA sample. A measurement is then performed on the transmitted field. In particular, we will discuss photon number and quadrature measurements and how they are impacted by the inevitable presence of losses (described by the loss rate $1-\eta$).
%We do not carry out the second squeezing operation and only consider losses after the OPA, where we drop the index in the remainder of this section. 

\subsection{Photon number measurements}

\begin{figure*}[]
\centering
\includegraphics[width=0.9\textwidth]{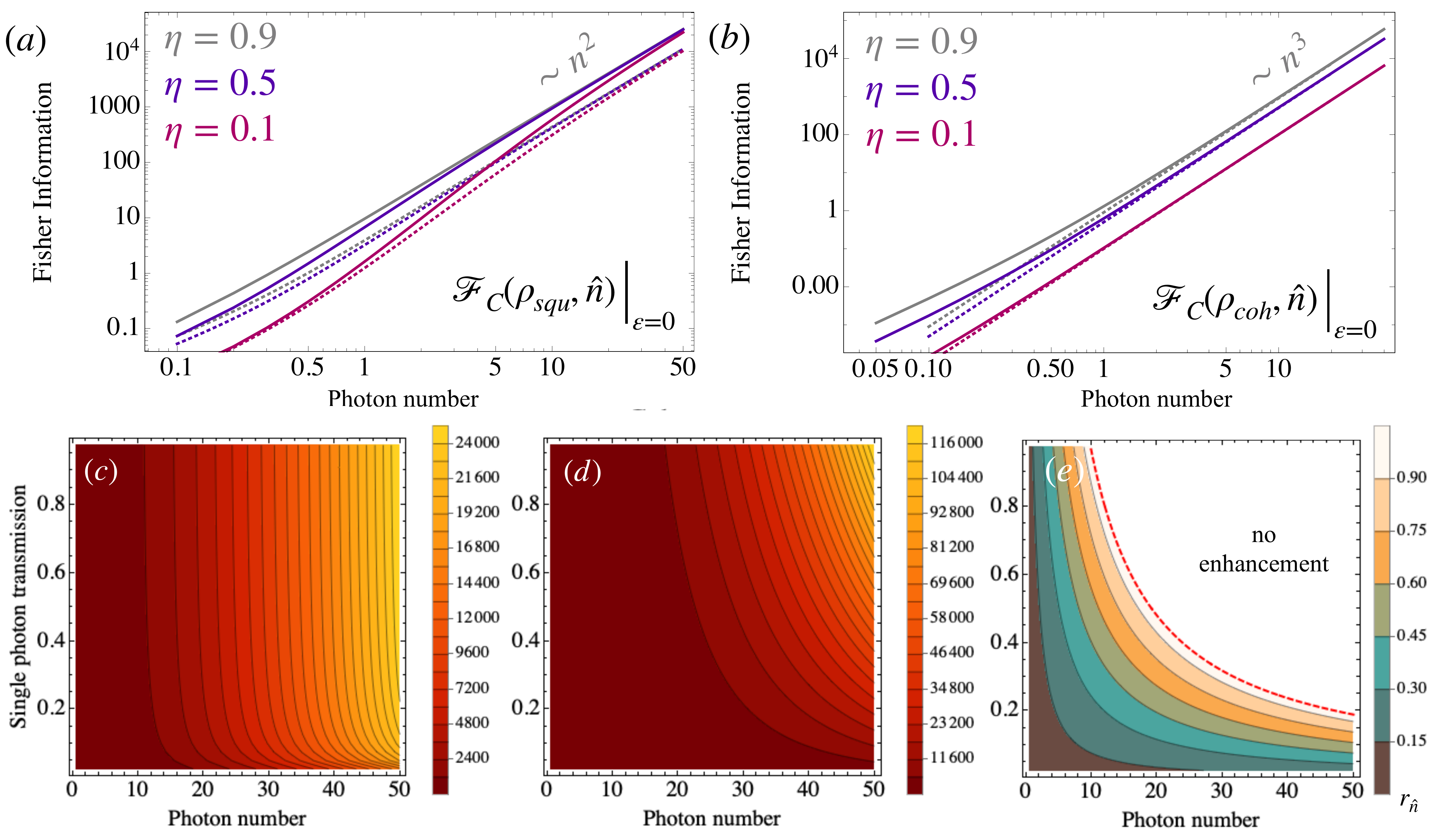}
\caption{ 
(a) Fisher information~(\ref{eq.FI_PhotoNum}) of photon number measurements is plotted vs. the mean photon number of squeezed vacuum input states in the presence of single photon losses with $\eta = 0.9$ (grey), $0.5$ (blue) and $0.1$ (red). The dashed lines correspond to the information attainable from the analysis of the mean photon number alone, i.e. the inverse of Eq.~(\ref{eq.DeltaEpsilon_n}). 
(b) Fisher information~(\ref{eq.FI_PhotoNum}) of photon number measurements is plotted vs. the mean photon number of coherent input states in the presence of single photon losses with $\eta = 0.9$ (grey), $0.5$ (blue) and $0.1$ (red). The dashed lines correspond to the information attainable from the analysis of the mean photon number alone according the inverse of Eq.~(\ref{eq.DeltaE_coh}). 
(c) Same as (a) plotted vs. the photon number and the single photon transmission probability $\eta$.
(d) Same as (b) plotted vs. the photon number and the single photon transmission probability $\eta$.
(e) Parameter regime, for which the ratio~(\ref{eq.r_n}) is below one, is plotted vs. the single photon transmission and the average photon number. The red dashed line indicates the crossover, where $r_{\hat{n}} = 1$. 
}
\label{fig.photonNumTPA}
\end{figure*}

\subsubsection{Squeezed vacuum}
We calculate the mean photon number, i.e. the expectation value of the operator $\mathcal{O} = a^\dagger a$, and obtain
\begin{align}
\langle \hat{n} \rangle_{sv} &= \eta \sinh^2 (r) - \frac{\varepsilon \eta}{2} \left( 3 \cosh (2r) -1 \right) \sinh^2(r) \notag \\
&= \eta n_r \left( 1 - \varepsilon (1 + 3 n_r) \right), \label{eq.intensity}
\end{align}
where we used $n_r = \sinh^2 (r)$ and the subscript ''sv" indicates the squeezed vacuum. 
The first term is the transmission of the empty setup (without the TPA sample) and the second term, which is proportional to $\varepsilon$, is calculated using Eq.~(\ref{eq.O_f}).
As expected, in the imperfect detection case the TPA and linear scattering losses cannot be separated. The measurement of the change of transmission can accurately and straightforwardly measure TPA cross sections in the presence of linear losses in the sample only if one can drive the absorption process in the quadratic regime $n_r \gg 1$.

Calculating further the variation of the photon number and using Eq.~(\ref{eq.sensitivity}), we obtain the sensitivity
\begin{align}
\Delta\varepsilon^{2 (sv)}_{\hat{n}} &= \frac{ 1 + \eta \cosh (2r) }{ \eta \sinh^2 (r) \left( 1 - 3 \cosh(2 r) \right)^2 /4} \notag \\
&= \frac{1}{ \eta n_r } \frac{1 + \eta (2 n_r + 1)}{\left( 1 + 3 n_r \right)^2}. \label{eq.DeltaEpsilon_n}
\end{align}
%where the superscript $(sv)$ indicates the squeezed vacuum. 
%Note that here we added the subscript "TPA" to distinguish this sensitivity from the one of an SU(1,1) interferometer, which we will discuss below. 
At large photon numbers, $i.e.$ $n_r \gg 1$, this result simplifies to
\begin{align}
\Delta\varepsilon^{2 (sv)}_{\hat{n}} &\rightarrow \frac{2}{ ( 3n_r)^2}, \label{eq.DeltaEps_n_limit}
\end{align}
which remarkably is \textit{independent of the linear losses} described by $\eta$. This behaviour is shown in Fig.~\ref{fig.photonNumTPA}(a), where the inverse of the sensitivity~(\ref{eq.DeltaEpsilon_n}) is plotted as dashed lines both for almost ideal setups (i.e. with transmission probability $\eta = .9$), and intermediate case ($\eta = 0.5$) and extremely lossy setups ($\eta = 0.1$). In any case, the sensitivity converges to the limit~(\ref{eq.DeltaEps_n_limit}) at large squeezing. The necessary photon number, for which this crossover takes place, will depend on the amount of single photon losses, i.e. it takes place when $2\eta n_r \gg 1$ in the numerator of Eq.~(\ref{eq.DeltaEpsilon_n}).
Finally, we remark again that if we additionally take single photon losses into account, which happen ahead of the TPA process, i.e. if the sample does not interact with a pure squeezed quantum state of light, then these losses do impact the sensitivity, of course.

\paragraph*{Fisher information}

Next, we focus on limits on achievable sensitivity via Fisher information according to the Cramer-Rao bound~(\ref{eq.Cramer-Rao}). 
%As described in section~\ref{sec.sensitivity}, the FI quantifies the information content to be gained from a measurement and limits the achievable sensitivity according to the Cramer-Rao bound~(\ref{eq.Cramer-Rao}). 
We already calculated the FI~(\ref{eq.FI_PhotoNum}) in \cite{Carlos2021} in the absence of single photon losses. Here, we analyze how the FI is affected by this error source.

The probability of detecting $n$ photons after they undergo single photon losses is given by~\cite{Avenhaus2008}
\begin{align}
P_n (\varepsilon) &= \sum_{m>n}
\begin{pmatrix}
m\\
n
\end{pmatrix}
\eta^n (1-\eta)^{m-n} P_m^{(0)} (\varepsilon), \label{eq.probabilities_losses}
\end{align}
where $P_m^{(0)}$ denotes the probability to find $m$ photons in the light field before the single photon losses occur. For a squeezed vacuum state with mean photon number $\overline{n}$, this probability is given by $P_m^{(0)} = (m!) \overline{n}^{m/2} / \big[ 2^{m/2} *((m/2)!)^2 (1+ \overline{n})^{(m+1)/2} \big]$ (for even $m$). 
The initial state, and hence its populations, undergoes TPA according to the master equation~(\ref{eq.Lindblad}), which yields the change of the population distribution,
\begin{align}
\frac{\partial P^{(0)}_n (\varepsilon)}{ \partial \varepsilon } &= \frac{1}{2} \bigg( (n+2)(n+1) P^{(0)}_{n+2} - n (n-1) P^{(0)}_n \bigg).
\end{align}
Together with Eq.~(\ref{eq.probabilities_losses}), this allows us to evaluate the Fisher information~(\ref{eq.FI_PhotoNum}) straightforwardly. 

The FI is shown as solid lines in Fig.~\ref{fig.photonNumTPA}(a) for the same parameters as the sensitivity~(\ref{eq.sensitivity}) discussed before, i.e. for single photon loss strength ranging between weak (photon transmission probability $\eta = 0.9$) and very strong losses ($\eta = 0.1$). As in our earlier analysis of the sensitivity attainable from mean photon number measurements in Fig.~\ref{fig.photonNumTPA}(a), we find that the FI converges to the "ideal" situation of vanishing single photon losses at large photon numbers. 
The convergence takes place on the same photon number range as for the sensitivity before. At each photon number, the FI is larger than the inverse of Eq.~(\ref{eq.DeltaEpsilon_n}), such that additional information can be gained from higher statistical moments of the photon number distribution. The scaling with photon numbers remains the same in both cases, $\propto \langle \hat{n}\rangle^2$.
The lack of dependence on the single photon losses is further illustrated in Fig.~\ref{fig.photonNumTPA}(c), where the same FI is plotted vs. both the photon number and the transmission probability $\eta$. At sufficiently large photon numbers, the dependence on losses can only be observed for very small $\eta$, i.e. very large photon loss probabilities. 

\subsubsection{Coherent states}
To compare the above result with a conventional TPA measurement with laser pulses, we now calculate the sensitivity achievable with transmission measurements of a coherent state with complex amplitude $\alpha$. 
Using the TPA transformation of the photon operators in Eq.~(\ref{eq.L-trafo}), we find $\partial \langle \hat{n} \rangle_{coh} / \partial \varepsilon = - \eta |\alpha|^4$.
Given the variance of the coherent state $\text{Var} (\hat{n}) = \eta | \alpha |^2 = \langle \hat{n} \rangle_{\varepsilon = 0}$, we find
\begin{align}
\Delta\varepsilon_{\hat{n}}^{2 \;(coh)} &= \frac{ \eta |\alpha|^2 }{ \eta^2 |\alpha|^8 } = \frac{1}{\eta |\alpha|^6} = \frac{1}{\eta n_\alpha^3}, \label{eq.DeltaE_coh}
\end{align}
where $n_\alpha = |\alpha|^2$.
Crucially, since both the variance and the change of the photon number are linear in $\eta$, the influence of linear losses can never be removed from the sensitivity. In contrast, the photon number variance of the squeezed state in Eq.~(\ref{eq.DeltaEpsilon_n}) contains linear and quadratic terms in $\eta$. At large squeezing, the latter dominate and the $\eta$-dependence cancels. 

The inverse sensitivity for coherent state measurements is plotted in Fig.~\ref{fig.photonNumTPA}(b) vs. the average photon number $|\alpha|^2$ for different single photon loss rates. In contrast to the squeezed state measurements in Fig.~\ref{fig.photonNumTPA}(a), the three dependencies are parallel at sufficiently large photon numbers, with the single photon losses accounting for a constant reduction of the sensitivity. Nevertheless, at large photon numbers coherent states can outperform squeezed vacuum states even in the presence of strong single photon losses due to the superior, cubic scaling with the photon number, $\propto \langle \hat{n}\rangle^3$. Fig.~\ref{fig.photonNumTPA}(d) further illustrates that this remains true for any single photon loss rate and any photon number. In contrast to the squeezed vacuum in Fig.~\ref{fig.photonNumTPA}(c), the equipotential lines do not tend to become vertical.

\paragraph*{Fisher Information}

\begin{figure}[t!]
\centering
\includegraphics[width=0.45\textwidth]{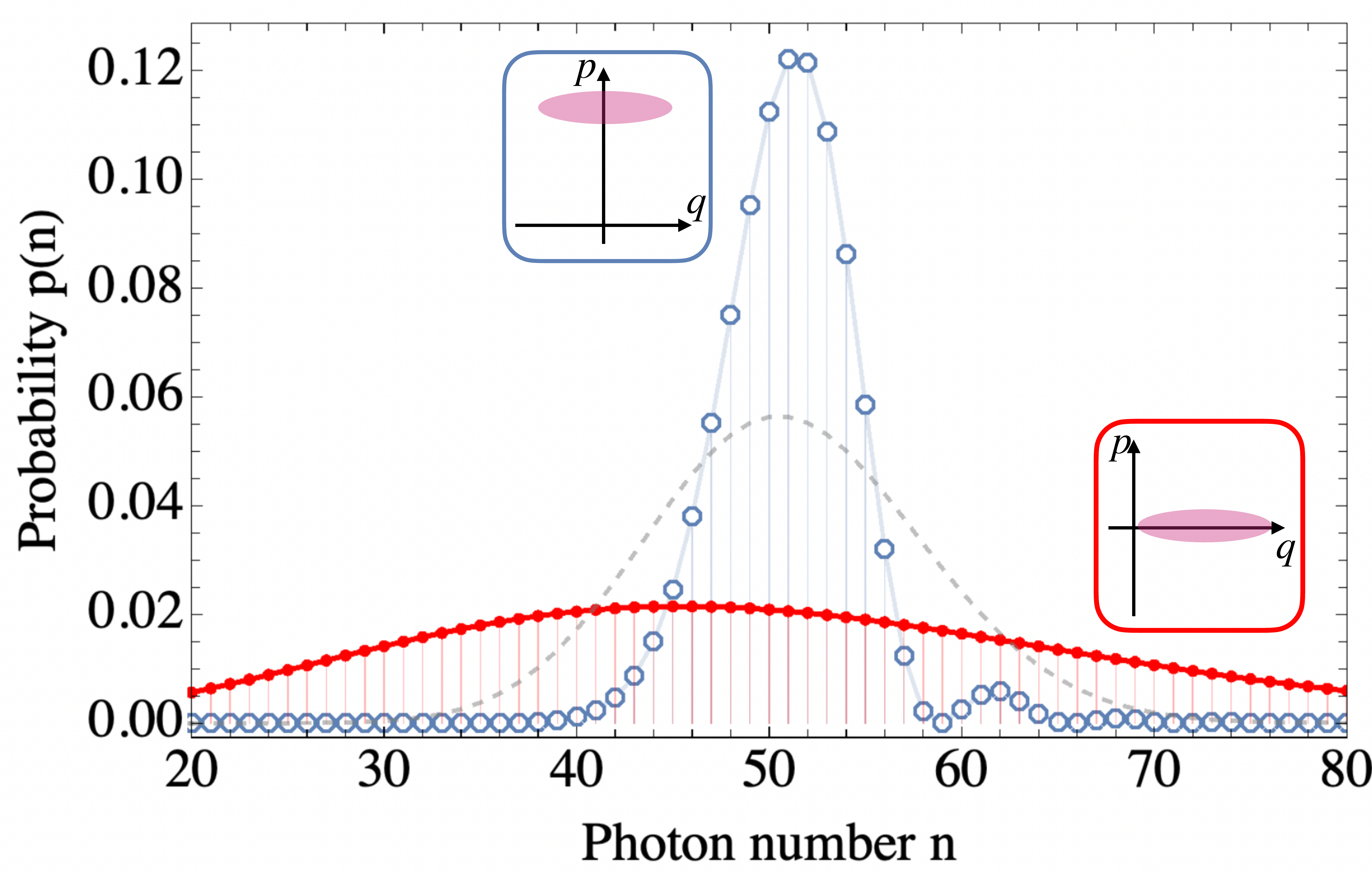}
\caption{ 
Photon number probability distributions of squeezed coherent states with fixed mean photon number $\langle \hat{n} \rangle = 50$. The open, blue circles show the probability distribution of an amplitude-squeezed state, where the laser phase is set to $\phi = \pi / 2$. The red points show the corresponding phase-squeezed states with $\phi = 0$. The insets sketch a phase space picture of the respective states, the frames are coloured as the respective plots. 
The gray, dashed line finally indicates a coherent state distribution for comparison. 
}
\label{fig.squeezed-coherent}
\end{figure}

\begin{figure}[h]
\centering
\includegraphics[trim=0 0 130 0, clip, width=0.48\textwidth]{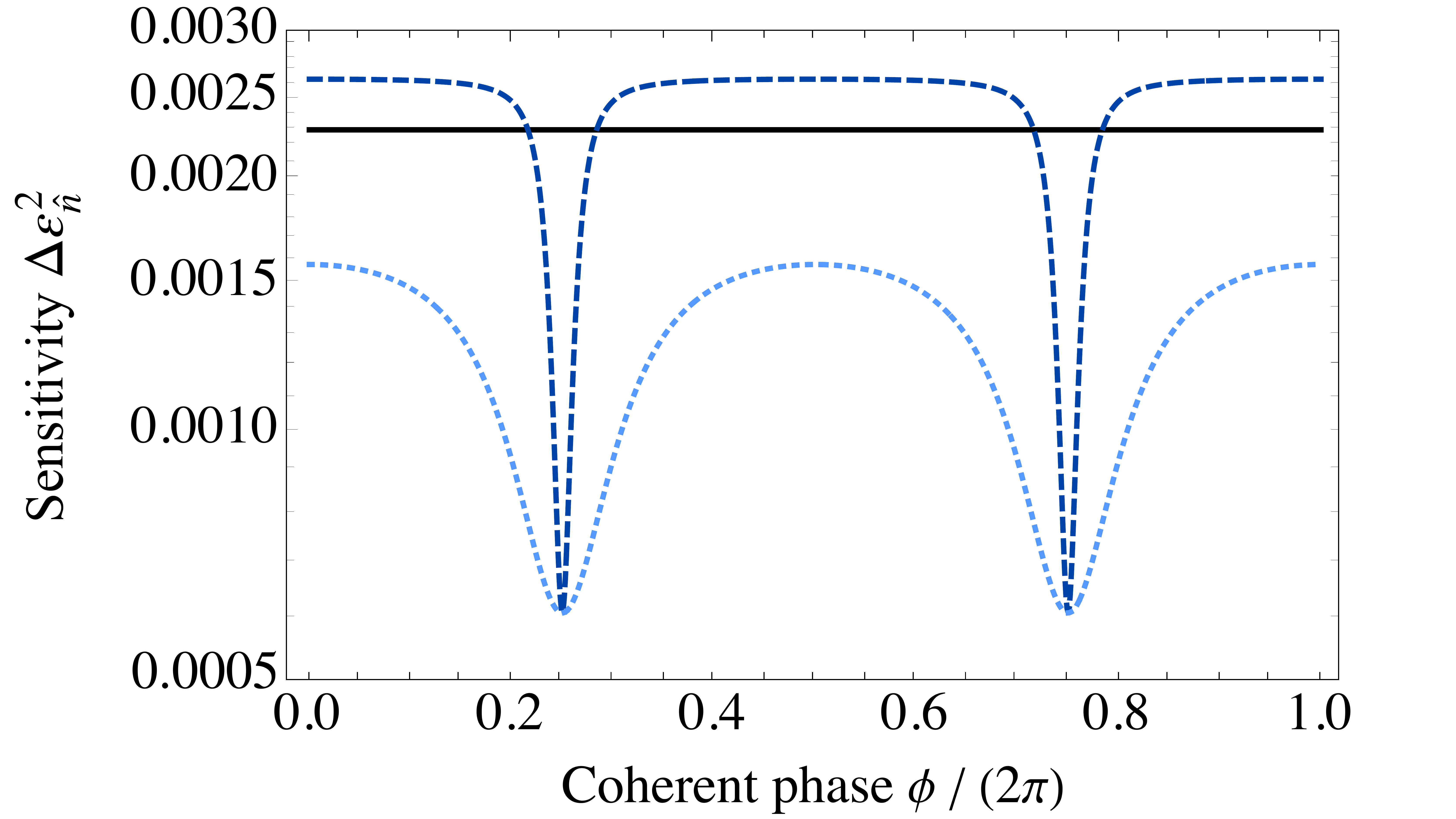}
\caption{ 
Phase dependence of the sensitivity~(\ref{eq.sensitivity}) for measurements with a squeezed coherent state with mean photon number $\langle \hat{n} \rangle = 10$ and squeezing parameters $r = 0.88$ (dashed, dark blue) and $r = 0.31$ (dotted, light blue). For each phase, we adjust the coherent amplitude $\alpha$ such that the mean photon number remains constant. 
The solid black line indicates the squeezed vacuum sensitivity at $\langle \hat{n} \rangle = 10$. We further set $\eta = 1$, i.e. we consider an ideal setup without any single photon losses.
}
\label{fig.phase-dependence_TPA}
\end{figure}

As in the squeezed vacuum case, we evaluate the FI~(\ref{eq.FI_PhotoNum}) which quantifies the additional precision gains that the analysis of the full photon counting distribution enables. We can analyze the calculations of the previous subsection using the probability distribution of a coherent state as input, i.e. we evaluate Eq.~(\ref{eq.probabilities_losses}) with $P_m^{(0)} = \exp (- \overline{n}) \overline{n}^m / m!$.
The results are shown as solid lines in Fig.~\ref{fig.photonNumTPA}(b). In contrast to the squeezed vacuum case in Fig.~\ref{fig.photonNumTPA}(a), the FI of coherent states and the inverse of Eq.~(\ref{eq.DeltaE_coh}) almost coincide for photon numbers $n \gtrsim 1$.
Hence, as one would intuitively expect, the analysis of higher statistical moments of the measured photon distribution cannot enhance the achievable sensitivity for TPA measurements with strong coherent light, where the probability distribution is dominated by its mean value and shows only small fluctuations. Furthermore, just like in Eq.~(\ref{eq.DeltaE_coh}), we find that single photon losses cannot be compensated and lower the FI linearly. 

We summarize the comparison between squeezed vacuum and coherent states in Fig.~\ref{fig.photonNumTPA}(e), where we plot the ratio
\begin{align}
r_{\hat{n}} \equiv \frac{ \mathcal{F}_C (\rho_{coh}, \hat{n})  }{ \mathcal{F}_C (\rho_{squ}, \hat{n})  } \label{eq.r_n}
\end{align}
between the FI of two case. Fig.~\ref{fig.photonNumTPA} shows the parameter regime, for which squeezed light can outperform coherent states, i.e. where $r_{\hat{n}} < 1$. This is the case at small photon numbers or in the presence of heavy single photon losses.
We find that when single photon losses are weak, i.e. $1 - \eta \simeq 0$, this is the case for $n\lesssim 10$.

\subsubsection{Squeezed coherent state}

Finally, we consider TPA detection with a squeezed coherent state, i.e. we assume that a coherent state with amplitude $\alpha$ is fed into the OPA as sketched in Fig.~\ref{fig.entTPA}. This setup creates the state $S (\zeta) D (\alpha) \vert 0 \rangle$. We note that this state is different from the usual definition of a squeezed coherent state $\vert \alpha, \zeta \rangle = D (\alpha) S (\zeta) \vert 0 \rangle$ which is often found in the literature, and where a squeezed vacuum is displaced. We can transform from on into the other using the identity
\begin{align}
\vert \alpha, \zeta \rangle &= S (\zeta) D (\alpha \cosh r + \alpha^\ast e^{i \phi} \sinh r) \vert 0 \rangle, \label{eq.squcoh-state}
\end{align}
and use, for instance, the photon number distributions for $\vert \alpha, \zeta \rangle$ in~\cite{BarnettRadmore}. 
With a general phase, $\alpha = |\alpha| e^{i \phi}$, we obtain a phase-squeezed coherent state at $\phi = 0$ and an amplitude-squeezed state at $\phi = \pi/2$. The photon number distributions of the two cases is shown in Fig.~\ref{fig.squeezed-coherent}. 

The average detected photon number is given by
\begin{widetext}
\begin{align}
\langle \hat{n} \rangle_{sq-coh} &= \eta \left( n_r + |\alpha|^2 \left( 1 + 2 n_r + \cos (2\phi) 2 \sqrt{ n_r (1+n_r) } \right) \right) \notag \\
&+ \varepsilon \eta \bigg[ \left( n_r + 3 n_r^2 \right) + 2 |\alpha|^2 \left( 2 n_r (2 + 3 n_r) + \cos (2\phi) \sqrt{n_2 (1+n_r)} (1 + 6 n_r) \right) \notag \\
&+ |\alpha|^4 \left( 1 + 6 n_r (1+n_r) + 4 \cos (2\phi) \sqrt{n_r (1+n_r)} (1 + 2 n_r) + 2 \cos(4\phi) n_r (1+n_r) \right) \bigg]. \label{eq.n_sc}
\end{align}
We remind the reader that we have used the abbreviation $n_r = \sinh^2 (r)$. The subscript ''sq-coh" distinguishes it from the squeezed vacuum calculation in Eq.~(\ref{eq.intensity}). 
The photon number variance at $\varepsilon = 0$ is given by
\begin{align}
\text{Var} (\hat{n}) &= \eta n_r \left( 1 + \eta (1 + 2 n_r) \right) \notag \\
&+ \eta |\alpha|^2 \left[ 1 + 2 n_r + \cos (2 \phi) \sqrt{n_r (1 + n_r)} + 2\eta \left( n_r (3 + 4 n_r) + \cos (2\phi) \sqrt{n_r (1+n_r)} (1 + 4 n_r) \right) \right].
\end{align}
\end{widetext}
The phase dependence of the resulting sensitivity is plotted in Fig.~\ref{fig.phase-dependence_TPA}. We fix the mean photon number that interacts with the sample, it is simply given by Eq.~(\ref{eq.n_sc}) at $\varepsilon = 0$ and $\eta = 1$ (i.e. before any losses have taken place). Overall, we find that amplitude-squeezed state states with $\phi = \pi / 2$ perform better than phase-squeezed states at $\phi = 0$. As we chose a fairly large mean photon number $\langle \hat{n} \rangle_{sq-coh} = 10$ and no single-photon losses, a large coherent amplitude is beneficial due to its cubic scaling, see Eq.~(\ref{eq.DeltaE_coh}), such that a weakly squeezed state generally performs better than a more strongly squeezed state or a squeezed vacuum state.
In any case, however, we find that amplitude-squeezing can provide an advantage, while phase-squeezing appears to be the least effective way to increase the sensitivity. 

\begin{figure}[]
\centering
\includegraphics[width=0.49\textwidth]{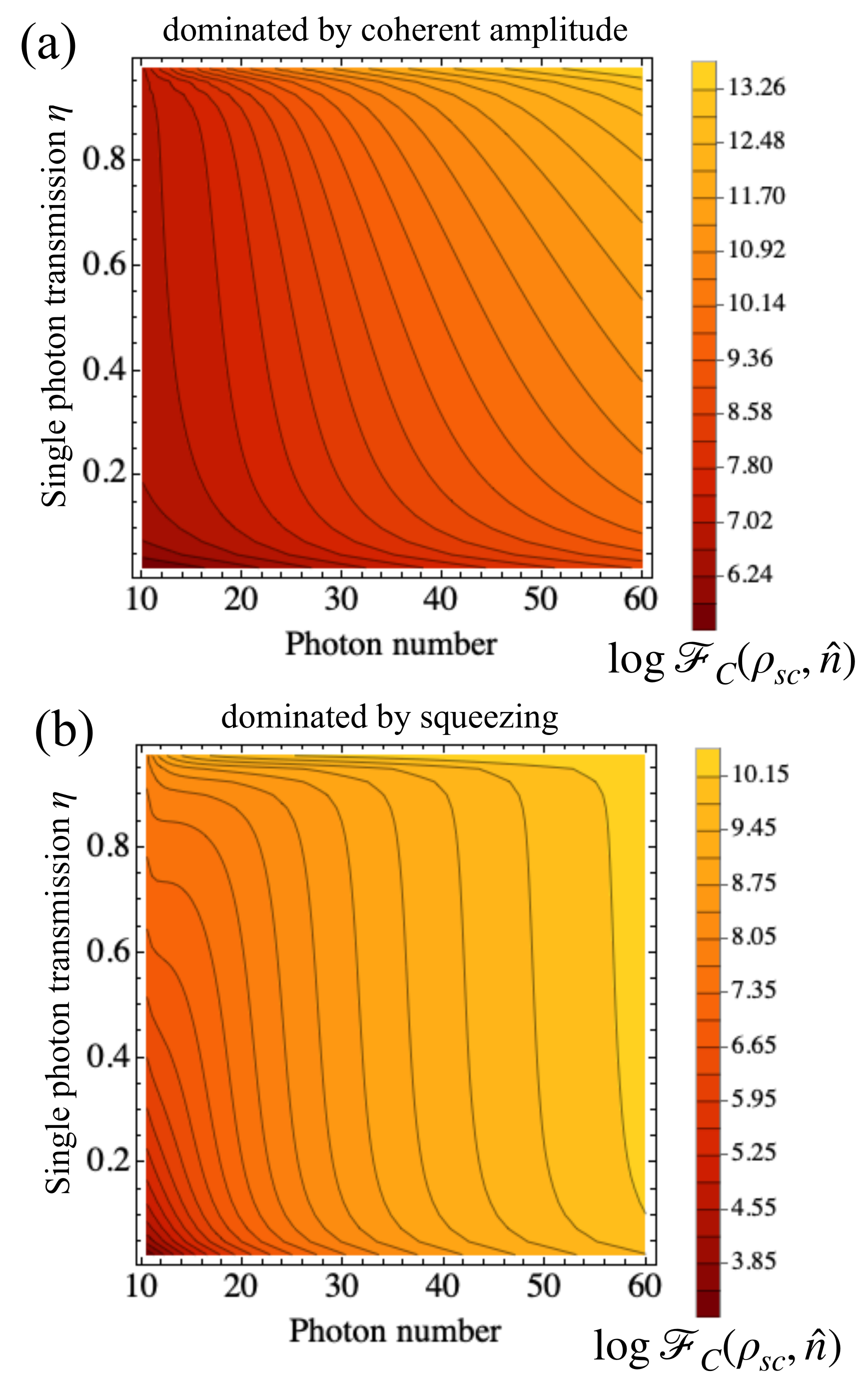}
\caption{ 
(a) Logarithm of the Fisher Information~(\ref{eq.FI_PhotoNum}) for an amplitude-squeezed coherent state (i.e. $\phi = \pi/2$), which is dominated by its coherent amplitude, is shown vs. the transmission probability $\eta$ and the average photon number. The squeezing parameter is fixed as $r = 1.876$ (corresponding to a photon number $\simeq 10$ in a squeezed vacuum state), and the coherent amplitude is increased to produce the desired average photon number. 
(b) Same as (a), but for a state with fixed coherent amplitude $\alpha = \sqrt{10}$ and variable squeezing parameter. 
}
\label{fig.SquCoh-PhotonNum}
\end{figure}

At large squeezing $n_r \gg 1$, the leading contributions give, using the definition of the sensitivity~(\ref{eq.sensitivity}), 
\begin{align}
\Delta\varepsilon^{2 (sc)}_{ \hat{n}} &\rightarrow \frac{ 2 \eta^2 |\alpha|^2 \left( 4n_r^2 + \cos (2\phi) 4 n_r^2 \right) }{ ( \eta |\alpha|^4 \left( 6n_r^2 + 8 n_r^2 \cos(2\phi) + 2 n_r^2 \cos (4 \phi) \right) )^2 } \notag \\
&= \frac{ 4 (1 + \cos(2\phi)) }{ |\alpha|^6 n_r^2 (16 \cos^4 \phi)^2 }.
\end{align}
At $\phi = 0$, i.e. a phase-squeezed state, this simplifies to
\begin{align}
\Delta\varepsilon^2_{\hat{n}, p-s} &\simeq \frac{1 }{ |\alpha|^6 e^{4r}} = \frac{1}{ |\alpha|^2 |\alpha e^{r}|^4 }. \label{eq.DeltaEpsilon-PhaseSqu-limit}
\end{align}
The subscript ''p-s" is introduced to denote the phase-squeezing.
Hence, the squeezed coherent state still shows the cubic scaling of the coherent input (i.e. $\sim |\alpha|^6$). But a sufficiently strong squeezing component can compensate for the detrimental influence of single photon losses, and render the sensitivity independent of the transmission probability $\eta$.
For an amplitude-squeezed state with $\phi = \pi / 2$, we find that both leading-order terms in numerator and denominator vanish, the subleading contributions then yield the sensitivity
\begin{align}
\Delta\varepsilon^2_{\hat{n}, a-s} &\simeq \frac{1 }{ |\alpha|^6 e^{2r}}. \label{eq.DeltaEpsilon-AmpSqu-limit}
\end{align}
Thus, at a first glance, it appears that the phase-squeezed state, Eq.~(\ref{eq.DeltaEpsilon-PhaseSqu-limit}) is capable of detecting TPA losses with higher precision. However, this changes completely when we relate the scaling to the incident photon number (Eq.~(\ref{eq.n_sc}) at $\varepsilon = 0$ and $\eta =1$), where we find $\langle \hat{n} \rangle_{p-s} \simeq | \alpha e^r |^2$ and $\langle \hat{n} \rangle_{a-s} \simeq | \alpha e^{-r} |^2$, respectively, at large $\alpha$. Here, we have neglected the spontaneously down-converted photons $\sim \sinh^2 r$. 
Plugging these back into the sensitivities~(\ref{eq.DeltaEpsilon-PhaseSqu-limit}) and (\ref{eq.DeltaEpsilon-AmpSqu-limit}), we find $\Delta\varepsilon^2_{\hat{n}, p-s} \simeq e^{2r} \langle \hat{n} \rangle^{-3}$ and $\Delta\varepsilon^2_{\hat{n}, a-s} \simeq e^{-4r} \langle \hat{n} \rangle^{-3}$. 
The squeezing thus deteriorates the sensing capabilities of the phase-squeezed state. The increase in the mean photon number is more than counter-balanced by the increased photon number fluctuations (compare Fig.~\ref{fig.squeezed-coherent}). The opposite is true for the amplitude-squeezed state: While the squeezing operation reduces the initial number of photons in the coherent state, $\alpha \rightarrow \alpha e^{-r}$, the resulting state shows exponentially enhanced sensitivity at a given average photon number. 
%This behaviour is exemplified in Fig.~\ref{fig.SquCoh-PhotonNum}(a). In the limit of large seeding and strong squeezing, the sensitivity approaches the limit~(\ref{eq.DeltaEpsilon-PhaseSqu-limit}), which is independent of single photon losses. 

\paragraph*{Fisher Information}
We finish this section by evaluating the FI~(\ref{eq.FI_PhotoNum}) in the presence of single-photon losses with squeezed coherent input states. In Figs.~\ref{fig.SquCoh-PhotonNum}(a) and (b) we present parameter scans of the FI vs. the transmission probability $\eta$ and the average photon number for two extreme cases: we keep the squeezing parameter~$r$ fixed in Fig.~\ref{fig.SquCoh-PhotonNum}(a), producing a state with substantial squeezing but whose large-$n$ behaviour is dominated by the coherent amplitude $\alpha$. Conversely, we fix this coherent amplitude in Fig.~\ref{fig.SquCoh-PhotonNum}(b) and increase the squeezing parameter~$r$. We find that the former states are much more strongly affected by single-photon losses. Nevertheless, as they show a cubic scaling at large photon numbers $\propto n^3$, the absolute FI can still become substantially larger than in the corresponding squeezing-dominated case. At fixed photon number, an optimal state will therefore contain the minimal amount of squeezing, which should be chosen sufficiently large to counter single-photon losses, but not larger than this. This makes it possible to benefit from the positive properties of the squeezed vacuum state, while also benefiting from the superior photon number scaling of the coherent state.

\begin{figure}[]
\centering
\includegraphics[width=0.45\textwidth]{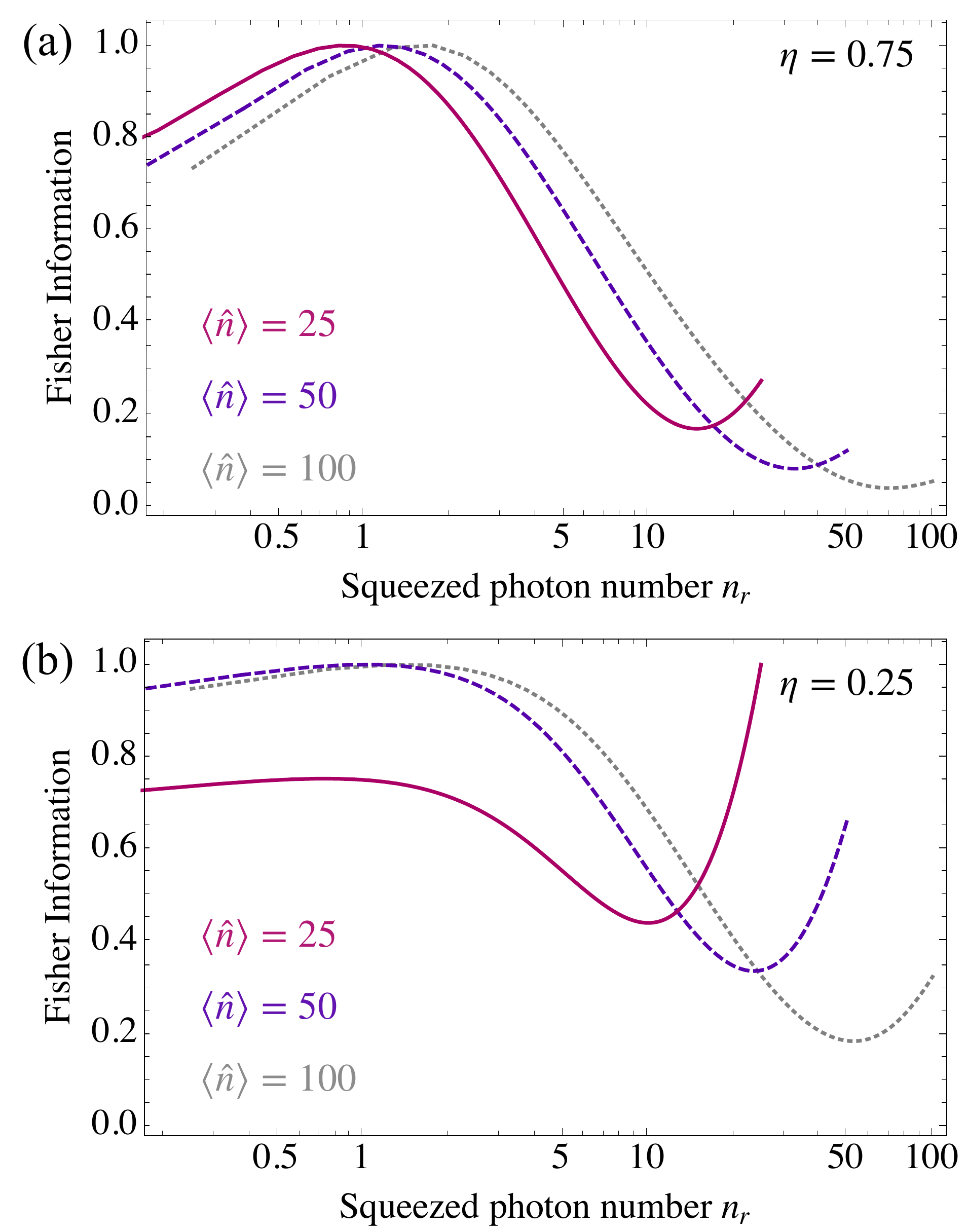}
\caption{ 
(a) The Fisher Information~(\ref{eq.FI_PhotoNum}), normalized to its optimal value, is plotted vs. the ''squeezed photon number" $n_r = \sinh^2 (r)$ at fixed single-photon transmission probability $\eta = 0.75$ and average photon number $\langle \hat{n} \rangle = 25$ (red), $\langle \hat{n} \rangle = 50$ (blue), and $\langle \hat{n} \rangle = 100$ (gray). 
We also fix the coherent amplitude phase $\phi = \pi / 2$, i.e. we focus on the amplitude-squeezed states that proved superior in Fig.~\ref{fig.phase-dependence_TPA}.
(b) same as (a) for $\eta = 0.25$.
}
\label{fig.optimal}
\end{figure}
This is illustrated in Fig.~\ref{fig.optimal}, where the FI of an amplitude-squeezed state is plotted vs. the squeezed photon number $n_r = \sinh^2 (r)$ at fixed total average photon number. 
This means with increasing $r$, the coherent amplitude is reduced to keep $\langle \hat{n} \rangle$ fixed. Each plot then terminates at a maximal squeezing parameter when the coherent amplitude vanishes, and $\langle \hat{n} \rangle  = \sinh^2 (r_{\text{max}})$. 
Furthermore, to compare different photon numbers, we normalize each plot to the maximal FI. 
In Fig.~\ref{fig.optimal}(a), simulations are shown for fairly weak single-photon losses $1-\eta = 0.25$. We find that indeed the FI increases first at small $r$, reaches its maximum and then drops to a minimum, before starting to increase again when the squeezed contribution becomes dominant. When the single photon losses are increased in Fig.~\ref{fig.optimal}(b), this increase at strong squeezing becomes more pronounced until it crosses the local maximum at smaller squeezing parameters. This is the case for $\langle \hat{n} \rangle = 25$ in Fig.~\ref{fig.optimal}(b), which means that the optimal state is a squeezed vacuum and we have reached the phase space region shown in Fig.~\ref{fig.photonNumTPA}(e), where squeezed states outperform coherent states. 

%\textcolor{red}{Make figure: FI vs. squeezing at fixed photon number}

\subsection{Quadrature measurements}

We next turn to measurements of the field position $q = (a + a^\dagger)/\sqrt{2}$ and the momentum quadrature $p = (a - a^\dagger)/(\sqrt{2}i)$. We remind the reader that in Eq.~(\ref{eq.OPA1}) we set the phase of squeezing $\phi_r = 0$. This means that, if squeezing is present, the quadrature $\hat{p}$ will become squeezed, while $\hat{q}$ will be anti-squeezed.

\subsubsection{Squeezed vacuum}

For a squeezed vacuum state, the expectation value of either quadrature vanishes, i.e. $\langle \hat{q} \rangle = \langle \hat{p} \rangle = 0$. Consequently, the sensitivity according to Eq.~(\ref{eq.sensitivity}) diverges, and in order to assess the metrological power of quadrature measurements in the presence of noise, we need to turn to the FI. 

\paragraph*{Fisher information}

%We conclude the section with an analysis of the Fisher information. 
In the presence of losses into an auxiliary mode $c$, we can use the transformation~(\ref{eq.loss-trafo}) to write the detected quadrature as
\begin{align}
\hat{q}_{detect} &= \sqrt{ \eta } \hat{q} + \sqrt{ 1 - \eta } \hat{q}_c, \label{eq.quadrature-mixing}
\end{align}
where we added the vacuum quadrature $\hat{q}_c = (c + c^\dagger) / \sqrt{2}$, which gives rise to a probability distribution $P_c (q)$ induced by these losses. As the auxiliary mode is in the vacuum, we have simply $P_c (q) = ( 2 / \pi )^{1/2} \exp (- 2 q^2) $. 
Defining the random variables $A \equiv \sqrt{ \eta } \hat{q}$ and $B \equiv \sqrt{1 - \eta} \hat{q}_c$, we find the distribution of $\hat{q}_{detect}$ as the sum of two independent random variables, 
such that the resulting probability distribution is given by the sum of those of the two random variables,  
\begin{align}
P (q) &= \int_{-\infty}^{\infty} P_0 \left( \frac{q}{\sqrt{ \eta}} - \sqrt{ \frac{1- \eta}{\eta} } q' \right) P_c (q') \frac{dq'}{ \sqrt{\eta} }, \label{eq.CFI-quad}
\end{align}
where $P_0$ denotes the probability distribution in the absence of losses.
Thus, single photon losses lead to a rescaling of the detected quadrature distribution, i.e. $q\rightarrow \sqrt{\eta} q$, as well as a convolution with the Gaussian distribution of the vacuum mode $c$.  

\begin{figure}
\includegraphics[trim= 0 0 180 0, width=0.49\textwidth]{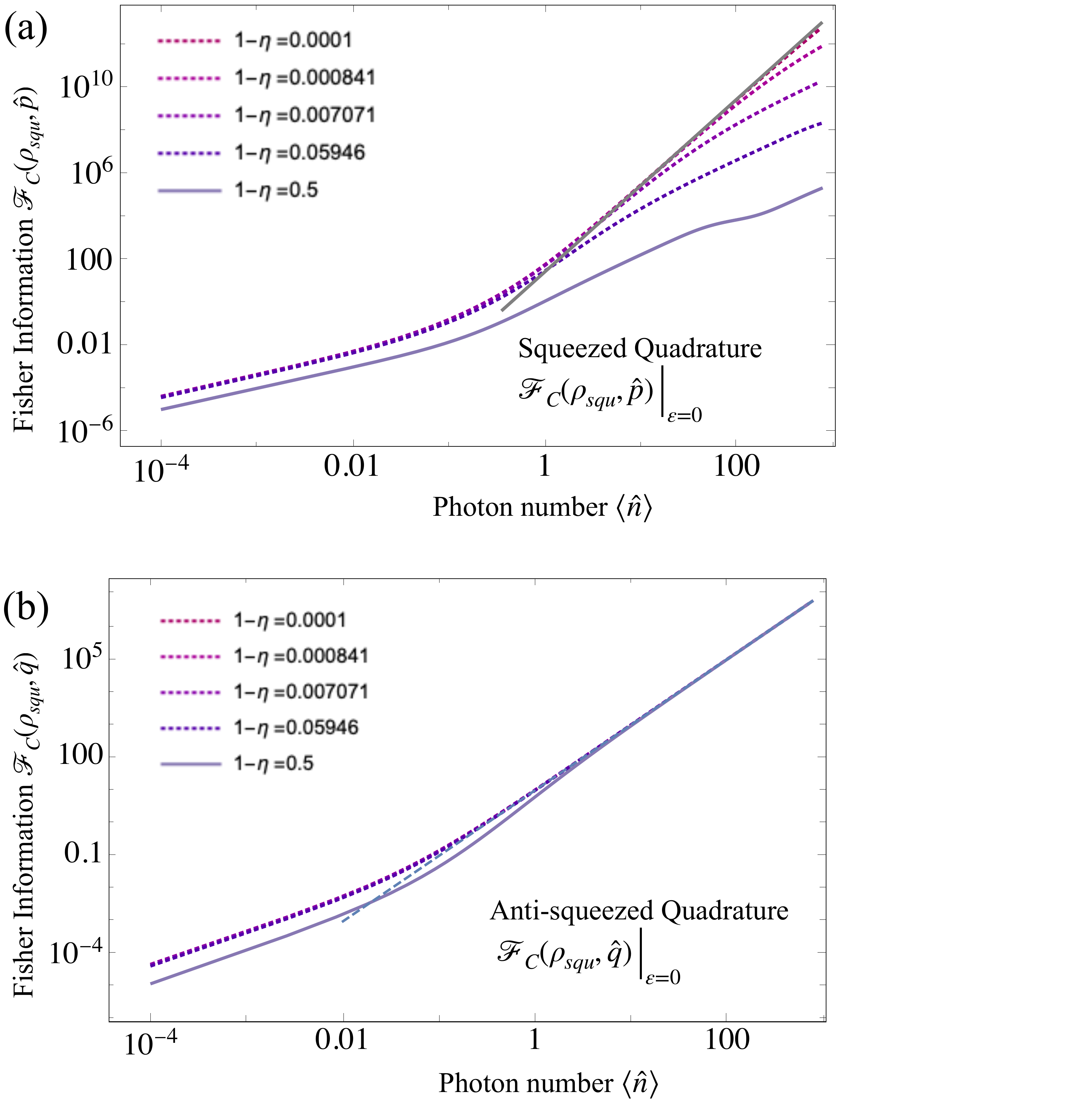}
\caption{ 
%\textcolor{red}{redo simulations for coherent state}
(a) Scaling of the Fisher Information of squeezed quadrature measurements, Eq.~(\ref{eq.CFI-quad}), for a squeezed vacuum input state is plotted vs. the average photon number. The gray line indicates the ideal scaling behaviour~(\ref{eq.FI-squeezed}).
(b) Same as (a), but for measurements of the anti-squeezed quadrature. The gray line indicates the corresponding scaling~(\ref{eq.FI-anti-squeezed}) of the FI at large photon numbers. 
}
\label{fig.quadratureTPA2}
\end{figure}

In \cite{Carlos2021}, we have calculated the FI for quadrature measurements of squeezed vacua in the absence of losses. In particular, we found a quartic scaling behaviour at large photon numbers for measurements of the squeezed quadrature, 
\begin{align}
\mathcal{F}_C^{0} (\rho_{squ}, \hat{p}) \sim 32 n^4, \label{eq.FI-squeezed}
\end{align}
and a quadratic scaling of the anti-squeezed quadrature,
\begin{align}
\mathcal{F}_C^{0} (\rho_{squ}, \hat{q}) \sim 21 n^2 /2. \label{eq.FI-anti-squeezed}
\end{align}
Note that in Eqs.~(\ref{eq.FI-squeezed}) and (\ref{eq.FI-anti-squeezed}), we only show the dominant scaling contribution at large photon number $n$. 
We investigate the degradation of these scaling behaviours by single photon losses in Fig.~\ref{fig.quadratureTPA2}. The FI of squeezed quadrature measurements is plotted in panel~(a). Evidently, even for tiny losses $1-\eta < 10^{-2}$ the quartic scaling~(\ref{eq.FI-squeezed}) is eroded substantially, and for $1-\eta = 0.5$ the scaling is lost entirely.
This is quite different for the anti-squeezed quadrature in panel~(b). For any loss rate $1-\eta$,  the FI eventually approaches the optimal scaling~(\ref{eq.FI-anti-squeezed}). As in our discussion of Fig.~\ref{fig.quadratureTPA} below, we find that measurements of the anti-squeezed quadrature are not affected by single-photon losses, provided the initial squeezing can overcome the loss rate.

\subsubsection{Coherent state}

\begin{figure}
\includegraphics[trim= 0 0 180 0, width=0.49\textwidth]{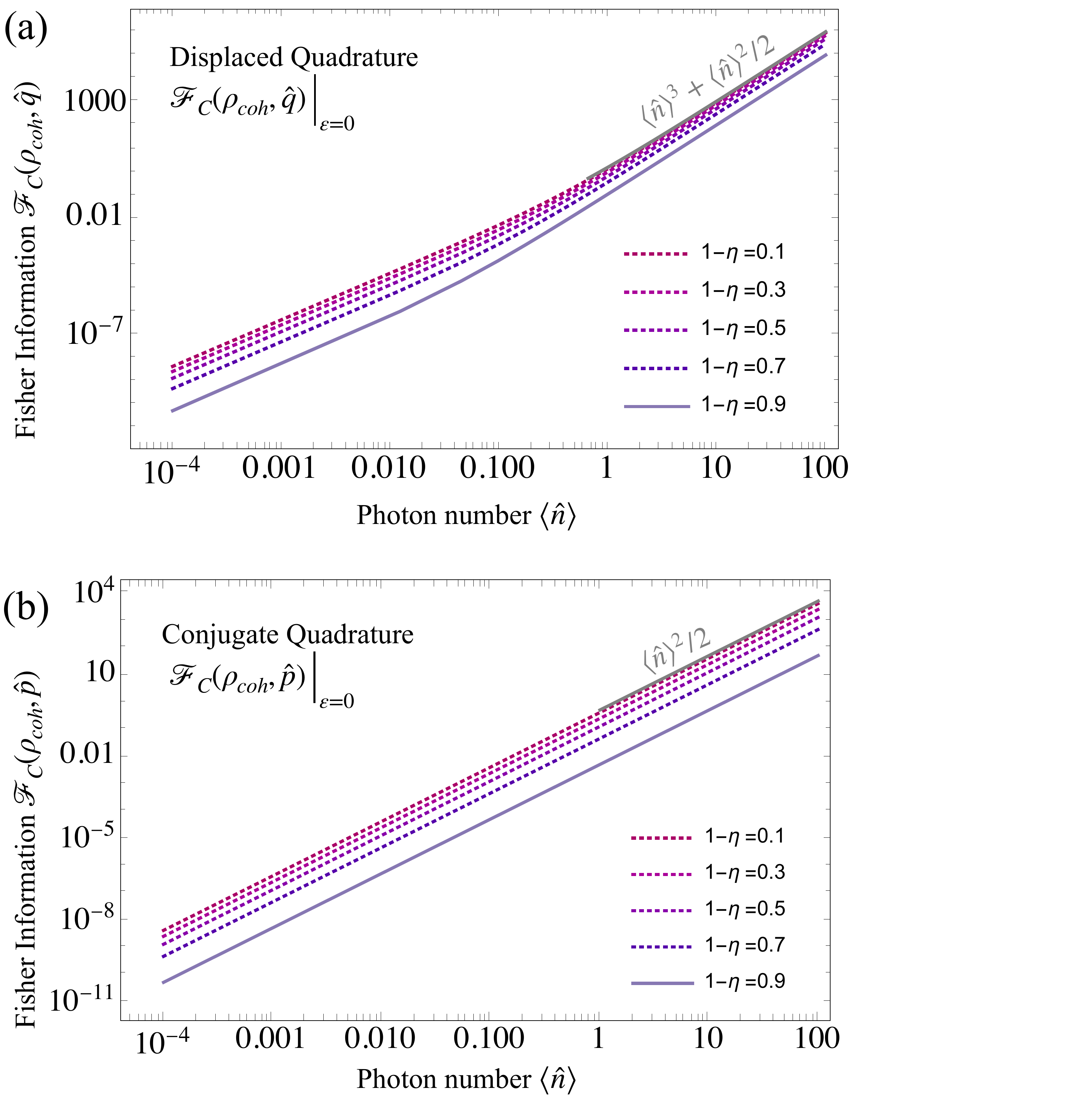}
\caption{ 
(a) Scaling of the Fisher Information of displaced quadrature measurements, Eq.~(\ref{eq.CFI-quad}), for a coherent input state is plotted vs. the average photon number. The gray line indicates the ideal scaling behaviour~(\ref{eq.F_CQ-coh}).
(b) Same as (a), but for measurements of the momentum quadrature. The gray line indicates the corresponding scaling~(\ref{eq.F_CP-coh}) of the FI at large photon numbers. 
}
\label{fig.quadratureTPA3}
\end{figure}

\begin{figure*}[]
\centering
\includegraphics[width=0.75\textwidth]{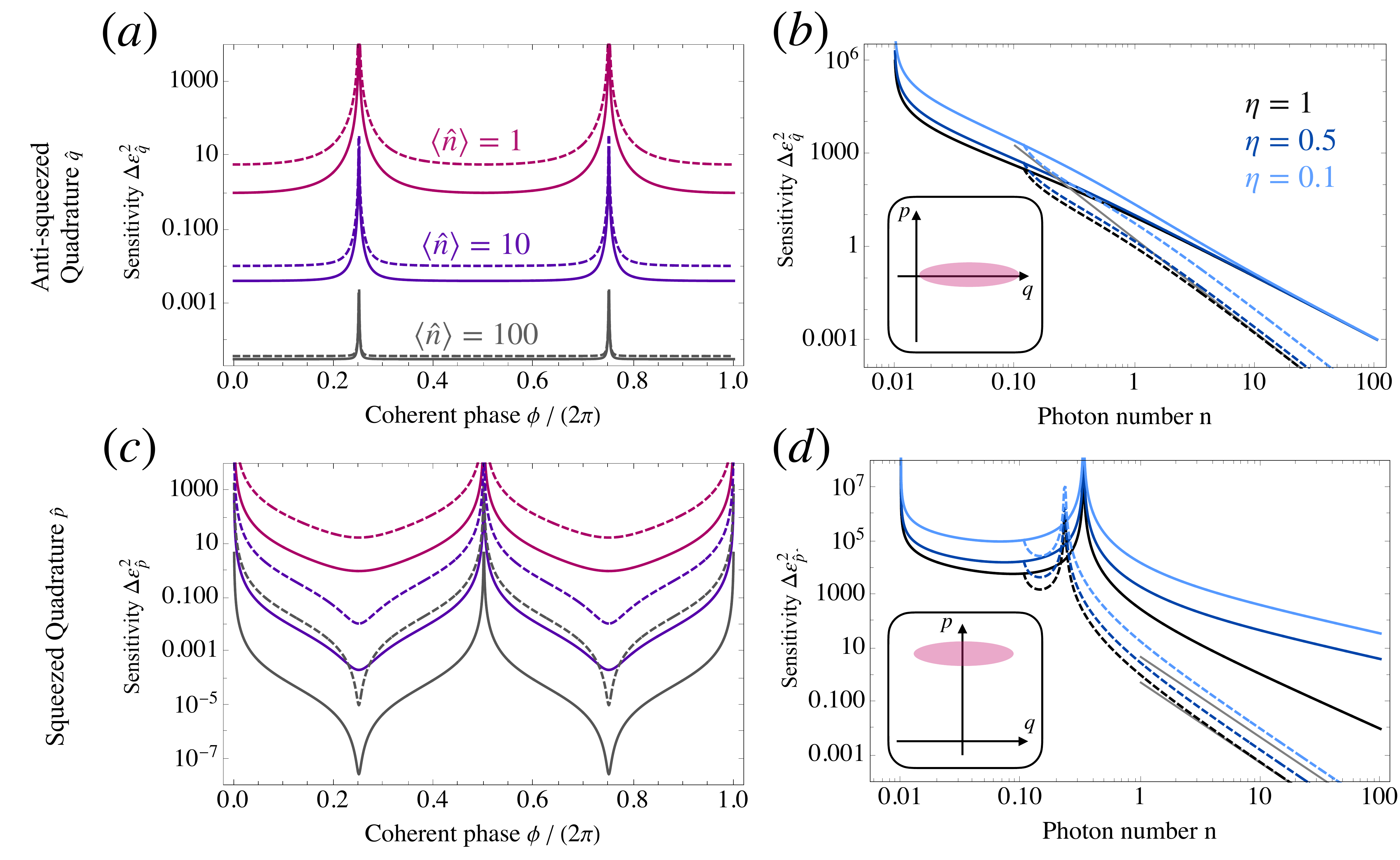}
\caption{ 
(a) Sensitivity $\Delta\varepsilon^{2}_{\hat{q}}$ for anti-squeezed $q$-quadrature measurements is plotted for different mean photon numbers vs. the coherent state phase $\phi$. The solid lines show an ideal measurement with $\eta= 1$, and the dashed lines a very noisy measurement with $\eta = 0.1$.
(b) Sensitivity $\Delta\varepsilon^{2}_{\hat{q}}$ is plotted vs. the mean photon number for different single photon transmission probabilities $\eta$. 
The solid lines correspond to squeezed coherent states, where the coherent amplitude is fixed at $\alpha = 0.1$ and the squeezing parameter $r$ is increased to generate larger photon numbers. The laser phase is fixed at $\phi = 0$. The light grey line indicates the limiting behaviour $e^{2r} / \langle \hat{n} \rangle^3$. The inset shows a sketch of the quadrature distribution of the incident phase-squeezed state.
(c) The same as (a) for squeezed quadrature measurements, $\Delta\varepsilon^{2}_{\hat{p}}$.
The dashed lines, in contrast, correspond to states where we fix $r = \sinh^{-1} (\sqrt{0.1})$ and increase the coherent amplitude $\alpha$. 
(d) The same as (b) for $\Delta\varepsilon^{2}_{\hat{p}}$. The light grey lines indicate the limiting behaviour $(1- \eta) \eta^{-1} e^{- 2r} \langle \hat{n} \rangle^{-3}$ for $\eta = 0.1$ and $1$. 
The laser phase is fixed at $\phi = \pi/2$, the inset again shows the corresponding phase space distribution. 
}
\label{fig.quadratureTPA}
\end{figure*}

For coherent input states, we obtain straightforwardly $\langle q \rangle^{(coh)} = \sqrt{2 \eta} |\alpha| \cos (\phi) (1 - \frac{\varepsilon}{2} |\alpha|^2)$ and $\langle p \rangle^{(coh)} = \sqrt{2\eta} |\alpha| \sin (\phi) (1 - \frac{\varepsilon}{2} |\alpha|^2)$, as well as $\text{Var} (p) = \text{Var} (q) = 1 / 2$. 
Hence, the sensitivity yields
\begin{align}
\Delta\varepsilon_{\hat{q}}^{2 \;(coh)} &= \frac{1}{\eta |\alpha|^6 \cos^2 (\phi)}, \\
\Delta\varepsilon_{\hat{p}}^{2 \;(coh)} &= \frac{1}{\eta |\alpha|^6 \sin^2 (\phi)}.
\end{align} 
In either case, quadrature measurements can only achieve the same sensitivity as photon number detection, Eq.~(\ref{eq.DeltaE_coh}), but never exceed it. They are further limited by single photon losses $\eta$ which cannot be removed or circumvented.

\paragraph*{Fisher information}

We can calculate the FI of coherent state measurements using the same approach as in the squeezed vacuum case. In \cite{Carlos2021}, we had derived the classical FI in an ideal setup. For a coherent state displaced along the $\hat{q}$-quadrature, we found
\begin{align}
\mathcal{F}_C^{0} (\rho_{coh}, \hat{q}) = n^3 + n^2 / 2, \label{eq.F_CQ-coh}
\end{align}
and
\begin{align}
\mathcal{F}_C^{0} (\rho_{coh}, \hat{p}) =  n^2 / 2. \label{eq.F_CP-coh}
\end{align}
The simulations in the presence of losses are shown in Fig.~\ref{fig.quadratureTPA3}. 
As in the photon number measurements, we find that the FI of both measurements is affected by single-photon losses. We never observe a convergence to the ideal setup, as we did in the anti-squeezed quadrature measurements before. However, it should also be noted that the degradation is not as dramatic as the reduction in Fig.~\ref{fig.quadratureTPA2}(a) of the FI of squeezed quadrature measurements of a squeezed vacuum.

\begin{figure*}[t]
\centering
\includegraphics[width=0.8\textwidth]{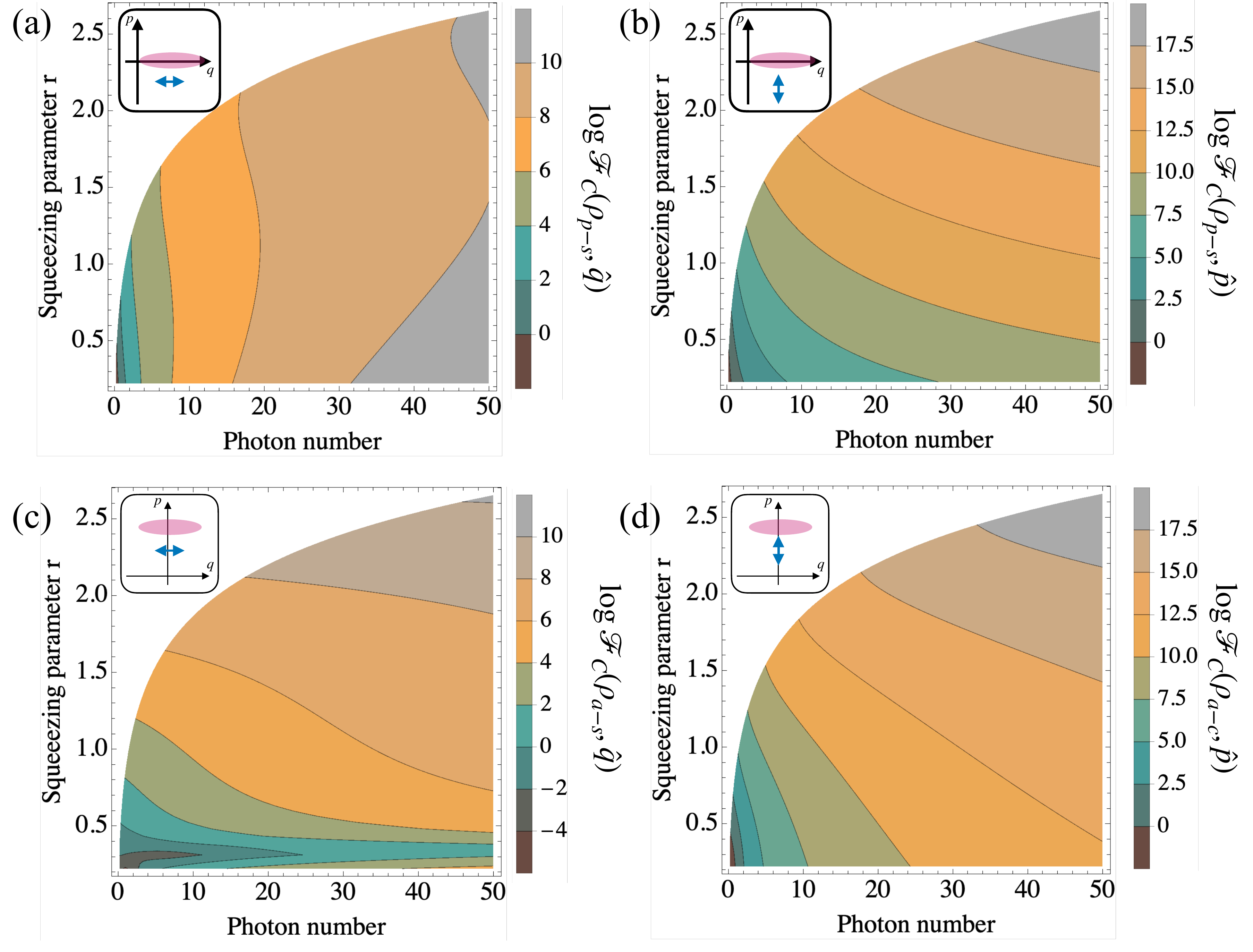}
\caption{ 
(a) Logarithm of the Fisher Information~(\ref{eq.quad-CFI}) for measurements of the anti-squeezed quadrature $\hat{q}$ for phase-squeezed state $\rho_{p-s}$ is shown vs. the average photon number of the incident state and the squeezing parameter $r$. Single-photon losses are not taken into account, i.e. we set $\eta = 1$. The inset shows an illustration of the state in phase space, the blue double-arrow indicates the quadrature measurement.
(b) Same as (a), but for measurements of the anti-squeezed quadrature $\hat{p}$.
(c) Same as (a) for measurements with an amplitude-squeezed state $\rho_{a-s}$. 
(d) Same as (b) for measurements with an amplitude-squeezed state. . 
}
\label{fig.SquCoh-quad}
\end{figure*}

\subsubsection{Squeezed coherent state}
We obtain for a squeezed coherent input state the expectation value
\begin{align}
\langle q \rangle &= \sqrt{2 \eta}\alpha \cos (\phi) e^{r} \notag \\
&- \frac{\varepsilon}{2} \sqrt{ \frac{\eta}{2}}\alpha \cos (\phi) e^r \bigg( 4 n_r + \sqrt{n_r (1+n_r)} + |\alpha|^2 f_r (\phi ) \bigg),
\end{align}
where we defined the factor $f_r (\phi ) = 1 + 2 n_r + \cos (2\phi) 2 \sqrt{ n_r (1+n_r) } = 2 \cosh (2r) +2 \cos(2\phi) \sinh(2r)$ which determines the seed amplification, and for optimal amplification at $\phi = 0$, i.e. for a phase-squeezed state, it evaluates simply to $f_r ( 0 ) = 2 \exp(2 r)$, and for an amplitude-squeezed state, $\phi = \pi/2$, it evaluates to $f_r (\pi/2) = 2 \exp(-2r)$.
The variance at $\varepsilon = 0$ yields
\begin{align}
\text{Var} (\hat{q}) &=\frac{1}{2} e^r \left( \sqrt{1+n_r} + \sqrt{n_r} \left( 2\eta -1 \right) \right). 
\end{align}
For measurements of the squeezed quadrature, we likewise obtain
\begin{align}
\langle p \rangle &= \sqrt{2 \eta}\alpha \sin (\phi) e^{-r} \notag \\
&- \frac{\varepsilon}{2} \sqrt{ \frac{\eta}{2}}\alpha \sin (\phi) e^{-r} \bigg( 4 n_r - \sqrt{n_r (1+n_r)} + |\alpha|^2 f_r (\phi ) \bigg), 
\label{eq.deriv-p}
\end{align}
and
\begin{align}
\text{Var} (\hat{p}) &=\frac{1}{2} e^{-r} \left( \sqrt{1+n_r} - \sqrt{n_r} \left( 2\eta -1 \right) \right).
\end{align}
These results are readily plugged into Eq.~(\ref{eq.sensitivity}) to obtain the sensitivity for measurements using the expectation value.
At large squeezing $n_r \simeq \exp (2r) /4 \gg 1$, the expression for the anti-squeezed quadrature simplifies to
\begin{align}
\Delta\varepsilon^2_{\hat{q}} &= \frac{ \text{Var} (\hat{q})  }{ | \partial \langle q \rangle/\partial\varepsilon |^2 } \label{eq.var-q} \\
&\simeq \frac{8 \eta \sqrt{n_r} e^r}{ \eta |\alpha \cos(\phi)|^2 e^{2r} ( 5 n_r + |\alpha|^2 f_r(\phi) )^2 } \notag \\
&= \frac{4}{|\alpha \cos(\phi)|^2 ( 5 n_r +  |\alpha|^2 f_r(\phi) )^2 }.
\end{align}
The sensitivity is maximized at $\phi = 0, \pi, \ldots$. 
As we have seen in the discussion of photon number measurements earlier, however, one has to be careful with the assessment, as the sensitivity is not yet normalized to the incident photon flux. Yet if we plot the sensitivity vs. the laser phase for fixed photon numbers in Fig.~\ref{fig.quadratureTPA}(a), we find that indeed $\phi = 0$, i.e. a phase-squeezed state, is the optimal coherent setting for the detection of the anti-squeezed quadrature. 
Further, we can encounter two limiting cases: when the coherent amplitude is very small, i.e. $\alpha \ll 1$, the sensitivity approaches $\Delta\varepsilon_q \rightarrow (25 |\alpha |^2 e^{4r} /16)^{-1}$. 
Translating this into a photon number dependence, this means $(25 |\alpha |^2 e^{4r} /16)^{-1} \simeq e^{- 2r} (25 \langle \hat{n} \rangle / 32)^{-1}$. 
In contrast, when the coherent seed dominates, $\alpha \gg e^r$, the sensitivity is given by $\Delta\varepsilon_q \rightarrow ( |\alpha|^6 e^{4r} )^{-1} = e^{2r} / \langle \hat{n} \rangle^3$. This reproduces the scaling behaviour of photon number measurements in Eq.~(\ref{eq.DeltaEpsilon-PhaseSqu-limit}).
These distinct scaling behaviours are shown in Figs.~\ref{fig.quadratureTPA}(b). The solid lines depict the sensitivity~(\ref{eq.var-q}) of states which are dominated by their squeezed contribution. Therefore, they basically reproduce the scaling behaviour of squeezed vacua in Fig.~\ref{fig.photonNumTPA}(a). 
If the squeezed coherent state is instead dominated by the coherent amplitude, we rather observe an inverse cubic scaling of the sensitivity. 
%These distinct scaling behaviours are shown in Figs.~\ref{fig.quadratureTPA}(a) and (b), respectively, where we show the sensitivity~(\ref{eq.sensitivity}) for anti-squeezed quadrature measurements vs. the squeezed photon number $n_2 = \sinh^2 (r)$.  Note that in Fig.~\ref{fig.quadratureTPA}(b), we fixed the coherent amplitude $\alpha = 100$, such that the total number of photons in the light field is much larger than in Fig.~\ref{fig.quadratureTPA}(a). 
%These distinct scaling behaviours are shown in Figs.~\ref{fig.quadratureTPA}(a), where we fix $\alpha = 0.1$. 
%Our presentation highlights the crossover from a low-squeezing regime, where single-photon losses impact the sensitivity to the limiting behaviours, which are independent thereof. This crossover depends only on the squeezing parameter $r$ (or equally on $n_r = \sinh^2 (r)$) and the loss rate $1 - \eta$, but not on the coherent amplitude.

Running the same analysis with the sensitivity of the squeezed $\hat{p}$-quadrature, we arrive at
\begin{align}
\Delta\varepsilon^2_{\hat{p}} &= \frac{ \text{Var} (\hat{p})  }{ | \partial \langle p \rangle/\partial\varepsilon |^2 }  \label{eq.var-p}\\
&=  \frac{8 e^{-r} (1-\eta) \sqrt{n_r}}{ \eta |\alpha \sin (\phi)|^2 e^{-2 r} \left( 3 n_r +  |\alpha|^2  f_r(\phi) \right)^2 } \notag \\
&\simeq \frac{1-\eta}{\eta} \frac{ 4 }{ |\alpha \sin (\phi)|^2  \left( 3 n_r + |\alpha|^2  f_r(\phi) \right)^2 }. 
\end{align}
Here we have approximated the numerator $\text{Var} (\hat{p}) \simeq (1-\eta) n_r^{1/2} \exp (r)$, which is valid for $n_r \gg 1$. 
First, we note that the impact of single-photon losses, described by the transmission factor $\eta$ (and loss rate $1-\eta$) cannot be eliminated from the expression. 
Furthermore, the sensitivity is largest when $\phi = \pi / 2, 3 \pi / 2, \ldots$, i.e. when we prepare an amplitude-squeezed state (see Fig.~\ref{fig.quadratureTPA}(c)). 
We again distinguish two limiting behaviours. For $\alpha \ll 1$, we find $\Delta\varepsilon^2_{\hat{p}} \simeq (1- \eta) \eta^{-1} 4 (3 \alpha e^{2r} / 4)^2 \simeq (1- \eta) \eta^{-1} 32 e^{-6r} / (9 \langle \hat{n} \rangle)$, where $\langle \hat{n}\rangle \simeq | \alpha e^{-r} |^2$. 
At large coherent amplitudes, we instead find $\Delta\varepsilon^2_{\hat{p}} \simeq (1- \eta) \eta^{-1} e^{- 2r} \langle \hat{n} \rangle^{-3}$. %This limiting behaviour is shown in Fig.~\ref{fig.quadratureTPA}(d).
%At these values, the second term in the squared expression vanishes, $e^{2r} + \cos (2\phi) e^{2r} = 0$, such that, for a strongly squeezed state with $n_r \simeq e^{2r} /4$,this expressions becomes $\Delta\varepsilon_p \eta/(1-\eta) = 16 (|\alpha e^r|^2)^{-1}$.
%When the coherent amplitude dominates, $\alpha \gg e^r$, we can optimize the sensitivity at $\phi = \pi /4$, where we find $\Delta\varepsilon_p \eta/(1-\eta) = 8 ( |\alpha|^6 e^{2r} )^{-1}$. 

We note that this scaling changes in an ideal measurement where $\eta = 1$. In this case, we obtain $\text{Var} (\hat{p}) = e^{-2r} /2$ (rather than the noisy limit $1-\eta$ used above), and an identical calculation shows that the sensitivity approaches (for a weakly seeded squeezed state with $\alpha \ll 1$) $\Delta\varepsilon^{(ideal)}_p \rightarrow 32 e^{- 6r} / (9 \langle \hat{n}\rangle)$. For a strongly seeded state with $\alpha \gg e^r$, we likewise obtain $\Delta\varepsilon^{(ideal)}_p (\phi = \pi /2) \rightarrow 2 e^{- 2r} / (\langle \hat{n}\rangle^3)$. However, the scaling remains inferior to the performance in photon number measurements, see Eq.~(\ref{eq.DeltaEpsilon-AmpSqu-limit}). 
%at the optimal laser phase $\phi = \pi /4$. This distinct scaling with the squeezing parameter $r$ can be seen in Fig.~\ref{fig.quadratureTPA}(c)-(d), where the ideal sensitivity is plotted in light blue. 

Eq.~(\ref{eq.var-p}) is plotted in Fig.~\ref{fig.quadratureTPA}(d) for different levels of single photon loss and for states with varying degrees of squeezing (solid lines) and varying coherent amplitudes (dashed lines). 
We observe a very interesting crossover between a low-photon regime, where the constant terms in the plot dominate, followed by a crossover regime where the coherent amplitude and the squeezing parameter give rise to similar photon numbers. Remarkably, in the course of this crossover the sensitivity diverges, since $\partial \langle \hat{p} \rangle / \partial\varepsilon$ changes sign when the $\alpha$-dependent term becomes larger than the second term in Eq.~(\ref{eq.deriv-p}). The approximate scaling behaviours are observed only at much larger mean photon numbers.

%Perhaps surprisingly, we find that the sensitivity of the anti-squeezed quadrature always shows superior scaling with respect to the squeezing parameter $r$. In addition, the sensitivity of $q$-measurements becomes independent of single-photon losses $\eta$. 

\paragraph*{Fisher information}

We finally simulate the FI of quadrature measurements with squeezed coherent states in Fig.~\ref{fig.SquCoh-quad}. 
For an ideal setup without single-photon losses, we present the FI of both quadrature measurements as a function of the average incident photon number and the squeezing parameter for phase- as well as amplitude-squeezed states. 
In our presentation, the horizontal line at $r = 0$ corresponds to the FI of coherent states. 
%This means, the line along the horizontal axis for $r = 0$ corresponds to the FI of coherent states shown in Fig.~\ref{fig.quadratureTPA3}. 
Conversely, the upper boundary, where we have maximal squeezing, corresponds to a squeezed vacuum, and the coherent amplitude vanishes, i.e. $\alpha = 0$. In between these two extreme cases, the relative strength of coherent and squeezed contributions is varied continuously. 

Measurements of the anti-squeezed quadrature are shown in Figs.~\ref{fig.SquCoh-quad}(a) and (c). In the former case, Fig.~\ref{fig.SquCoh-quad}(a), we find cubic growth of the FI according to Eq.~(\ref{eq.F_CQ-coh}), such that, at any fixed photon number, squeezing merely reduces the FI. In Fig.~\ref{fig.SquCoh-quad}(c), on the other hand, it grows quadratically according to Eq.~(\ref{eq.F_CP-coh}), such that the squeezed vacuum ultimately shows superior FI. We also note that, as in the case of the sensitivity in Eq.~(\ref{eq.var-p}), the FI goes through a local minimum as a function of the squeezing, before growing to the squeezed vacuum case. 

In Figs.~\ref{fig.SquCoh-quad}(b) and (d), we show the FI pertaining to measurements of the squeezed quadrature. Here, we find weaker differences between the phase-squeezed state in Fig.~\ref{fig.SquCoh-quad}(b) and its amplitude-squeezed counterpart in Fig.~\ref{fig.SquCoh-quad}(d). In both cases, the optimal state at large photon numbers is the squeezed vacuum due to its quartic increase according to Eq.~(\ref{eq.FI-squeezed}). 

%Measurements of the squeezed quadrature are simulated in Fig~\ref{fig.SquCoh-quad}(a). We find a very interesting, non-monotonic behaviour of the FI. There is a regime of weak squeezing, where the FI decreases with $r$. It then reaches a minimum and afterwards reaches a strong squeezing regime, where the FI becomes only dependent on $r$, and shows no dependence (or rather a very weak one) on the coherent amplitude anymore. The former regime is dominated by the coherent amplitude, and weak squeezing only comes at the cost of a reduction of this amplitude, thus reducing the FI. The latter regime, in contrast, is dominated by squeezing, which gives rise to the $\propto n_r^4$ scaling in the squeezed vacuum case, see Eq.~(\ref{eq.FI-squeezed}). This superior scaling can outperform a larger coherent amplitude, such that ultimately the FI only depends the squeezing parameter at large average photon numbers. 

%Measurements of the anti-squeezed quadrature are simulated in Fig~\ref{fig.SquCoh-quad}(b). In contrast to panel (a), the cubic scaling of the FI with the photon number, which we observe in coherent states, Eq.~(\ref{eq.F_CQ-coh}), always dominates. Consequently, the FI only shows a weak dependence on the squeezing parameter. 

\section{Conclusions}
\label{sec.conclusions}

\begin{table}[t]
  \centering
    \begin{tabular}{|c|c|c|c|}
    \hline
 \multirow{2}{*}{\textbf{Measurement}} &  \multirow{2}{*}{\textbf{State}} & \multicolumn{2}{c|}{\textbf{Sensitivity} $\Delta\epsilon^2$}\\ 
\cline{3-4}
& & Limit & Formula\\
\hline
&
Squeezed vacuum  &
\begin{minipage}[c][10mm][t]{0.1mm}%
\end{minipage} 
  $n_r \gg 1$ &\large $\frac{2}{9 n^2}$\\
   \cline{2-4}
\multirow{4}{*}{\textsc{Photon counting}}  &
Coherent state & N.A. &
\begin{minipage}[c][10mm][t]{0.1mm}%
\end{minipage}
 \large $\frac{1}{\eta n^3}$\\
   \cline{2-4}
& \thead{Squeezed coherent \\ $\phi = 0$ (Phase sq.)}& $n_r \gg 1$ &
\begin{minipage}[c][10mm][t]{0.1mm}%
\end{minipage}\(\displaystyle \frac{e^{2r}}{n^3}\)
\\
\cline{2-4}
&\thead{Squeezed coherent \\ $\phi =\pi/2$\\ (Amplitude sq.)} &  $n_r \gg 1$ &
\begin{minipage}[c][10mm][t]{0.1mm}%
\end{minipage}\(\displaystyle \frac{e^{-4r}}{n^3}\)
\\
\hline
%
%%======QUADRATURE MESUREMENTS-q=====%%
&
Coherent state &N.A. &\begin{minipage}[c][10mm][t]{0.1mm}%
\end{minipage} \(\displaystyle
\frac{1}{\eta n^3\cos^2\phi}\)\\
\cline{2-4}
 \multirow{2}{*}{\thead{\textsc{Quadrature Measurements}\\ Anti-squeezed $q$-quadrature }}& \multirow{2}{*}{ \thead{Squeezed coherent \\ $\phi = 0$ (Phase sq.)}}& $\alpha \ll 1$ &
\begin{minipage}[c][10mm][t]{0.1mm}%
\end{minipage}\(\displaystyle 
e^{-2r}\frac{32}{25 n}
\)
\\
\cline{3-4}
& & $\alpha \gg e^r$ &
\begin{minipage}[c][10mm][t]{0.1mm}%
\end{minipage}\(\displaystyle 
\frac{e^{2r}}{n^3}
\)
\\
\hline
%%======QUADRATURE MESUREMENTS `p=====%%
&
Coherent state  & 
N.A. &\begin{minipage}[c][10mm][t]{0.1mm}%
\end{minipage} \(\displaystyle \frac{1}{\eta n^3\sin^2\phi}\)\\
\cline{2-4}
  \multirow{2}{*}{\thead{\textsc{Quadrature Measurements}\\Squeezed $p$-quadrature }}& \multirow{2}{*}{ \thead{Squeezed coherent \\ $\phi = \pi/2$\\ (Amplitude sq.)}} &  $\alpha \ll 1$ &
\begin{minipage}[c][10mm][t]{0.1mm}%
\end{minipage}\(\displaystyle 
\frac{1-\eta}{\eta}e^{-6r}\frac{32}{9n}
\)
\\
\cline{3-4}
& & $\alpha \gg 1$ &
\begin{minipage}[c][10mm][t]{0.1mm}%
\end{minipage}\(\displaystyle 
\frac{1-\eta}{\eta}e^{-2r}\frac{1}{n^3}
\)
\\
\hline

  \end{tabular}
\caption{Summary of analytical results for sensitivities of different states and measurement approaches. }
\label{tab.summary}
  \end{table}

In summary, we have investigated the detection of two-photon absorption in the presence of competing single-photon losses, which could stem from scattering losses or imperfect photon detectors. Focussing on the regime of very weak TPA losses, we provided an extensive analysis of the sensitivities achievable with different observables, namely the photon number, and the two field quadratures. 
A summary of the scaling behaviours we derived is provided in Table~\ref{tab.summary}. Please note that in the quadrature measurement results, we only list phase-squeezed (amplitude-squeezed) states in measurements of the anti-squeezed (squeezed) quadrature. As the expectation value of the anti-squeezed (squeezed) quadrature vanishes for an amplitude-squeezed (phase-squeezed) state, the corresponding variance $\Delta\varepsilon^2$ diverges. 

For photon number measurements, our main result is that the TPA detection sensitivity measured with squeezed vacuum states can become independent of single-photon losses, provided the mean photon number of the incident state is sufficiently large. In this high-intensity regime, the change in the observable's expectation value due to single-photon losses is exactly cancelled by the corresponding change of the photon number variance. This cancellation persists also for the corresponding Fisher information, thus cancelling the detrimental effect of the single-photon losses. 
It does not occur in measurements with coherent probe states, where the sensitivity is always reduced by unwanted losses. However, the latter sensitivity shows a superior cubic scaling with the photon number, such that coherent probes will still perform better at very large photon fluxes. 
Moreover, we find that coherent state measurements are determined entirely by the field expectation values. The calculation of the FI shows that higher order correlation measurements do not improve the sensitivity. 
This is in contrast to squeezed probes, where the FI is always roughly a factor two larger than the sensitivity derived from the photon number expectation value.

In our analysis of quadrature measurements, we find that 
%This picture persists for quadrature measurements. M
measurements of the squeezed quadrature are strongly affected by additional losses. The beneficial quartic scaling of the FI alluded to in the introduction is quickly eroded even by very small single-photon losses. 
In contrast, the anti-squeezed quadrature fluctuations compensate for these losses and become independent thereof. However, as in the case of photon number measurements, this comes at the price of a suboptimal quadratic scaling with the mean photon number. 

Finally, we also investigated measurements with squeezed coherent states, where we found that suitable tuning of the parameters of the light allows to combine the positive aspects of both squeezed and coherent probes. In particular, in photon counting experiments, single-photon losses can be compensated, provided the squeezing is sufficiently large to counteract the losses. 
Too much squeezing, however, has a detrimental effect on the sensitivity by reducing the superior cubic scaling of coherent probes. The competition between these two effects lead us to define 
% To this end, we found 
optimal degrees of squeezing for a given mean photon number and noise level. 
This behaviour is true for both amplitude- and phase-squeezed states of light (see Table~\ref{tab.summary}). However, the latter states suffer from exponentially large photon number fluctuations, such that an amplitude-squeezed state is beneficial for TPA measurements. 
This is no longer true in quadrature measurements, where the squeezed quadrature is always affected by single photon losses.
%Additionally, the cubic scaling of the sensitivity with the coherent photon number can be maintained. 

\begin{acknowledgments} 
 
S. P. and F. S. acknowledges support from the Cluster of Excellence 'Advanced Imaging of Matter' of the Deutsche Forschungsgemeinschaft (DFG) - EXC 2056 - project ID 390715994. 
C. S. M. acknowledges that the project that gave rise to these results received the support of a fellowship from la Caixa Foundation (ID 100010434) and from the European Union Horizon 2020 research and innovation program under the Marie Skodowska-Curie Grant Agreement No. 847648, with fellowship code LCF/BQ/PI20/11760026, and financial support from the Proyecto Sin\'ergico CAM 2020 Y2020/TCS-6545 (NanoQuCo-CM).
%C. S. M. acknowledges that the project that gave rise to these results received the support of a fellowship from la Caixa Foundation (ID 100010434) and from the European Union's Horizon 2020 Research and Innovation Programme under the Marie Sklodowska-Curie Grant Agreement No. 47648, with fellowship code  LCF/BQ/PI20/11760026.

\end{acknowledgments}

\end{document}